\pdfoutput=1 %figs in istr-paper-pdf-figs, to compile: pdflatex istr-paper-pdf-04.tex >>>istr-paper-pdf-03.pdf
\documentclass{article} %{\epsfig{ subtituted for \includegraphics[ only, 41 figures

\usepackage[a4paper, total={6in, 10in}]{geometry}

\usepackage[percent]{overpic} %does not work
\usepackage{amsmath}
\usepackage{multirow} %for tables
\usepackage{fancyvrb} %to center verbatim

\usepackage{enumitem} %so the whole list is not indented
\usepackage{easylist}  
\usepackage{graphicx}							
\usepackage{amssymb}
\usepackage[dvipsnames]{xcolor}%to fix pdflatex error 
\usepackage{epsf}

\usepackage{epsfig}
\usepackage{url}
\usepackage{wrapfig}
\usepackage{float} %to force table/figure with [H]
\newenvironment{s-itemize}{
\begin{itemize}
  \setlength{\itemsep}{1pt}
  \setlength{\parskip}{0pt}
  \setlength{\parsep}{0pt}
}{\end{itemize}}

\usepackage{enumitem} %to insert paras in {s_enumerate} list and continue count with [resume]

\newenvironment{s_enumerate}{
\begin{enumerate}
  \setlength{\itemsep}{1pt}
  \setlength{\parskip}{0pt}
  \setlength{\parsep}{0pt}
}{\end{enumerate}}

\newenvironment{s_itemize}{
\begin{itemize}
  \setlength{\itemsep}{1pt}
  \setlength{\parskip}{0pt}
  \setlength{\parsep}{0pt}
}{\end{itemize}}

\newenvironment{blurb-small} %to set footnotesize with correct spacing
  {\par\small}
  {\par\addvspace{\bigskipamount}}

\newenvironment{blurb-foot} %to set footnotesize with correct spacing
  {\par\footnotesize}
  {\par\addvspace{\bigskipamount}}
\usepackage[bottom]{footmisc} %to force footnote on same page as indicated

\usepackage[plainpages=false,backref=page]{hyperref} %include backtrack from refs
\usepackage[hypcap=true]{caption}

\title{\bf The istr-graph: Interactive Visualisation\\of any Classic-Graph in DDLab}
\author{Andrew Wuensche%
\thanks{andy@ddlab.org,  \url{http://www.ddlab.org}}}%

\date{\small August 2026}	% Activate to display a given date or no date

\begin{document}

\maketitle

\vspace{-5ex}
%^^^^^^^^^^^^^^^^^^^^^^^^^^^^^^^^^^^^^^^^^^^^^^^^^^^^^^^^^^^^^^^^^^^^
\begin{abstract}  
  \noindent Any type of attractor basin (classic-graph) created in
  DDLab can now be visualised, manipulated, and deconstructed as a
  drag/drop ``interactive state transition graph'' (istr-graph). The
  new istr-graph applies to subtrees, single basins, the basin of
  attraction field, compression, and all other classic-graph
  parameters. This is an important update on the pre-existing
  ``interactive basin of attraction field graph'' (ibaf-graph)
  specific to just the complete uncompressed field, but the ibaf-graph is
  nevertheless retained for some of its unique attributes. These
  issues are discussed with a focus on the scope and implementation of
  the new istr-graph.
\end{abstract}

\begin{center}
  {\it keywords: DDLab, cellular automata, random Boolean networks,
    discrete dynamical networks, directed graphs, random maps,
    basins of attraction, state transition graphs, interactive visualisation.}
\end{center}

%^^^^^^^^^^^^^^^^^^^^^^^^^^^^^^^^^^^^^^^^^^^^^^^^^^^^^^^^^^^^^^^^^^^^^^^^^^^^
\section{Introduction} 
\label{Introduction}

A central purpose of the Discrete Dynamics Lab software
(DDLab)\cite{Wuensche-DDLab} is to compute and visualise attractor basins,
state transition graphs, of finite discrete dynamical networks
(DDN), from cellular automata (CA)\cite{wolfram2002,Wuensche92}, to
random Boolean networks (RBN)\cite{kauffman69,kauffman93,Wuensche94a},
to more general DDN\cite{EDD} up to and including random
maps\cite{wuensche97}\hspace{-.2ex}\cite[\hspace{-1ex}\footnotesize{\#29.8}]{EDD}\footnote{References
  to the book ``Exploring Discrete Dynamics''(EDD)\cite{EDD} usually
  include the relevant chapter or section. EDD and
  DDLab\cite{Wuensche-DDLab} are kept updated online at
  \url{http://www.ddlab.org} and mirror sites.}.
Attractor basins are significant in many areas\cite{wuensche10} including
self-organisation\cite{kauffman93,langton90,wuensche97,wuensche99},
memory\cite{wuensche96,wuensche2005}, gene regulation\cite{kauffman69,somogyi96,wuensche98a,wuensche2002}
and as mathematical objects in their own right.

\begin{figure}[b]/
  \vspace{-2ex}
  \begin{center} 
  \begin{minipage}[t]{.9\linewidth}  
   \includegraphics[viewport=231 160 716 560,width=.48\linewidth,clip]{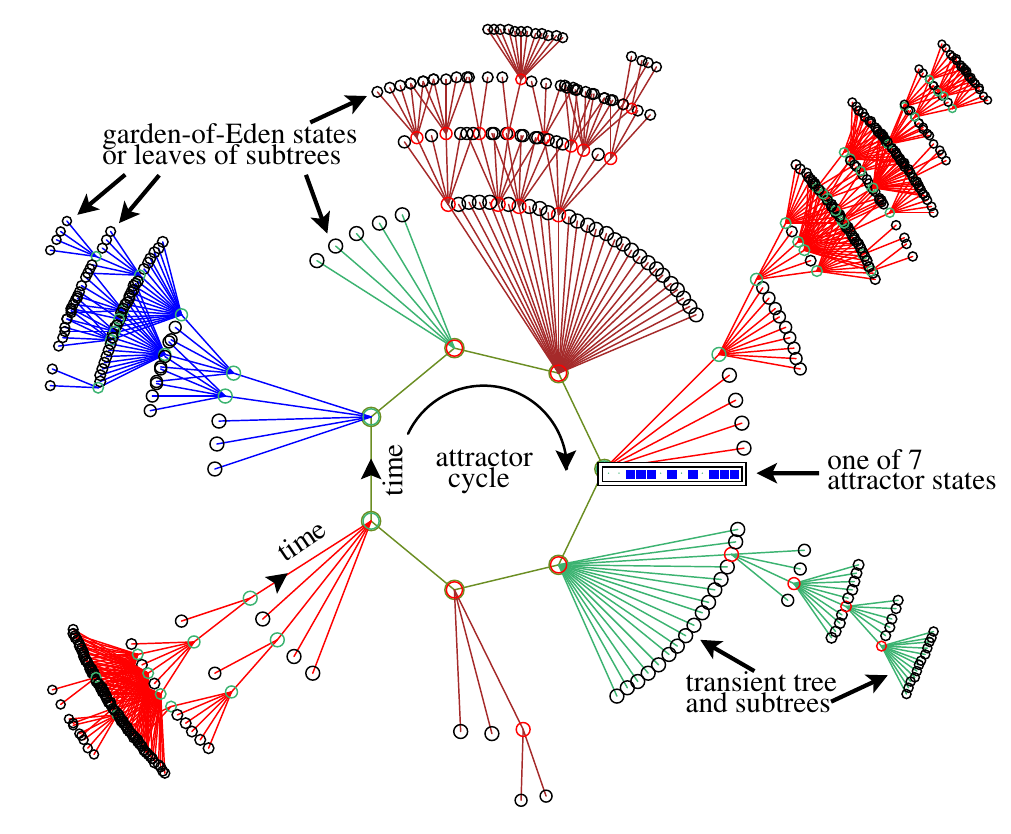}
   \hfill
   \includegraphics[viewport= 400 281 935 710, width=.5\linewidth]{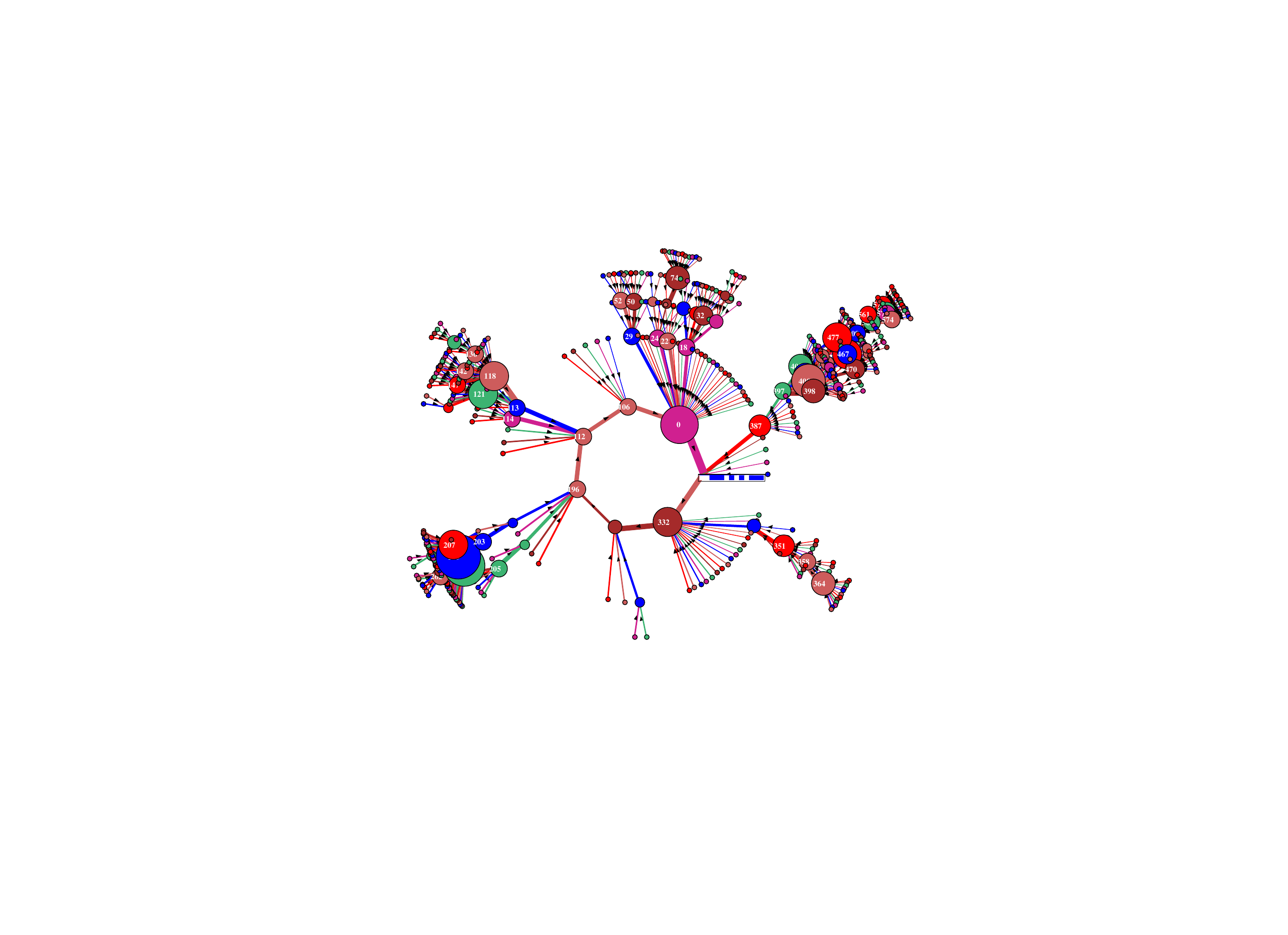}   
    \end{minipage}
    \end{center} %vector PostScript file %includes arrows,text 
   \vspace{-5ex}  
   \caption[A basin of attraction of a random Boolean network]
   {\textsf{
           A basin of attraction (one of 15 in the field)
           of a random Boolean network (RBN)  defined in 
               \mbox{\cite[\hspace{-1ex}\footnotesize{\#2.11.2}]{EDD}}.
           The basin has 604 states, of which 523 are
            leaf states.
               The direction of time is inwards
               from leaf states to the attractor, then clock-wise.
               An attractor state (dec=1879) is shown as a bit pattern.
               \underline{\it Left}: the {\it static} classic-graph with
               annotations added.
               \underline{\it Right}: the {\it interactive} istr-graph
               with both nodes (as discs) and edges scaled by inputs.
               Disc numbers are shown
               by computation order (if they fit).
     \label{fig:rbn_P}          
     }}
\label{fig:rbn_P.ps}      
\end{figure}

Attractors are found by forward iterations in discrete time-steps
until a repeat is encountered.  One of three reverse algorithms
computes the predecessors (pre-images) of each attractor state, then
the pre-images of pre-images and so on, to unravel transient trees backwards
in time, eventually reaching all leaf (garden-of-Eden) states. The
basin of attraction field (or a single basin) which may be subject to
``compression'' has the topology of trees rooted on cycles.% (figure~ \ref{r110ibaf.ps}).

\begin{figure}[t]
\vspace*{-3ex}
  \begin{center}
    \textsf{\small
    \begin{minipage}[t]{.85\linewidth}
      \includegraphics[width=1\linewidth,viewport=135 464 902 740,clip]{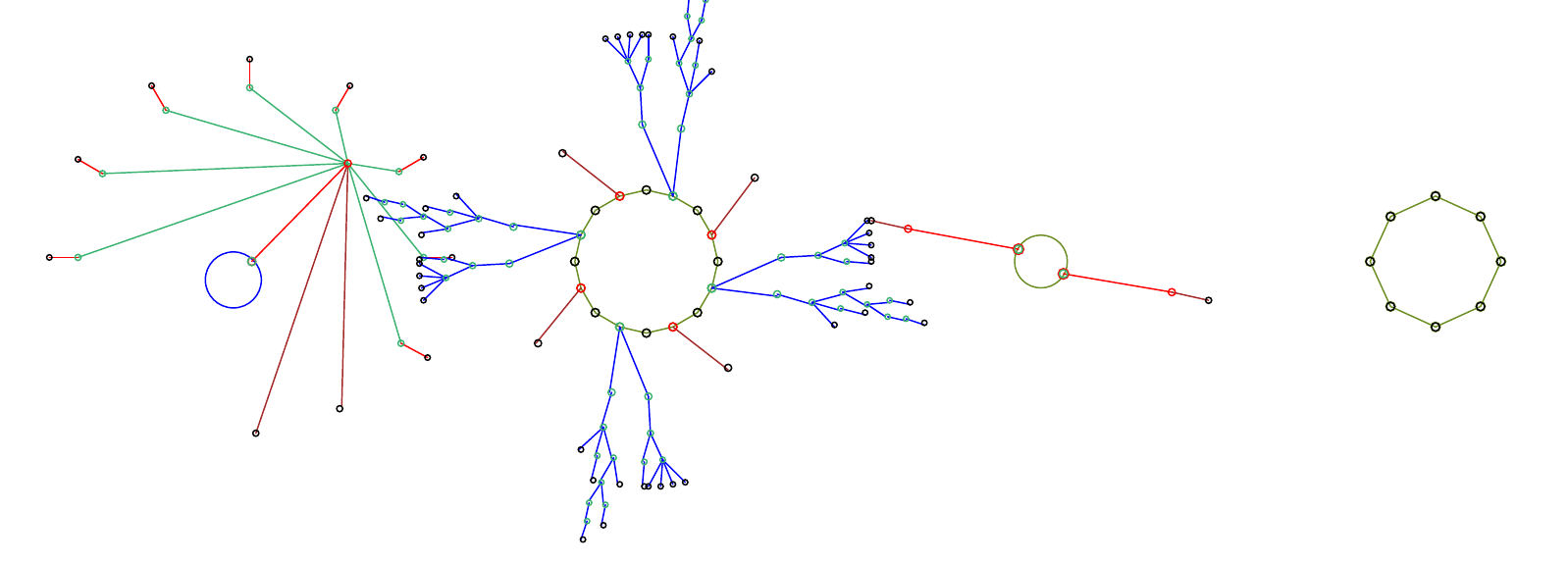}\\[-6ex]
     \begin{center}(a) classic-graph of the compressed basin of attraction field, default layout
     with 142 nodes.
     Although the layout can be preset before and during drawing,
     once complete the classic graph is static.\end{center}
     \end{minipage}\\[-.5ex]
   \begin{minipage}[t]{.85\linewidth}
     \includegraphics[width=1\linewidth,viewport=80 216 1187 665,clip]{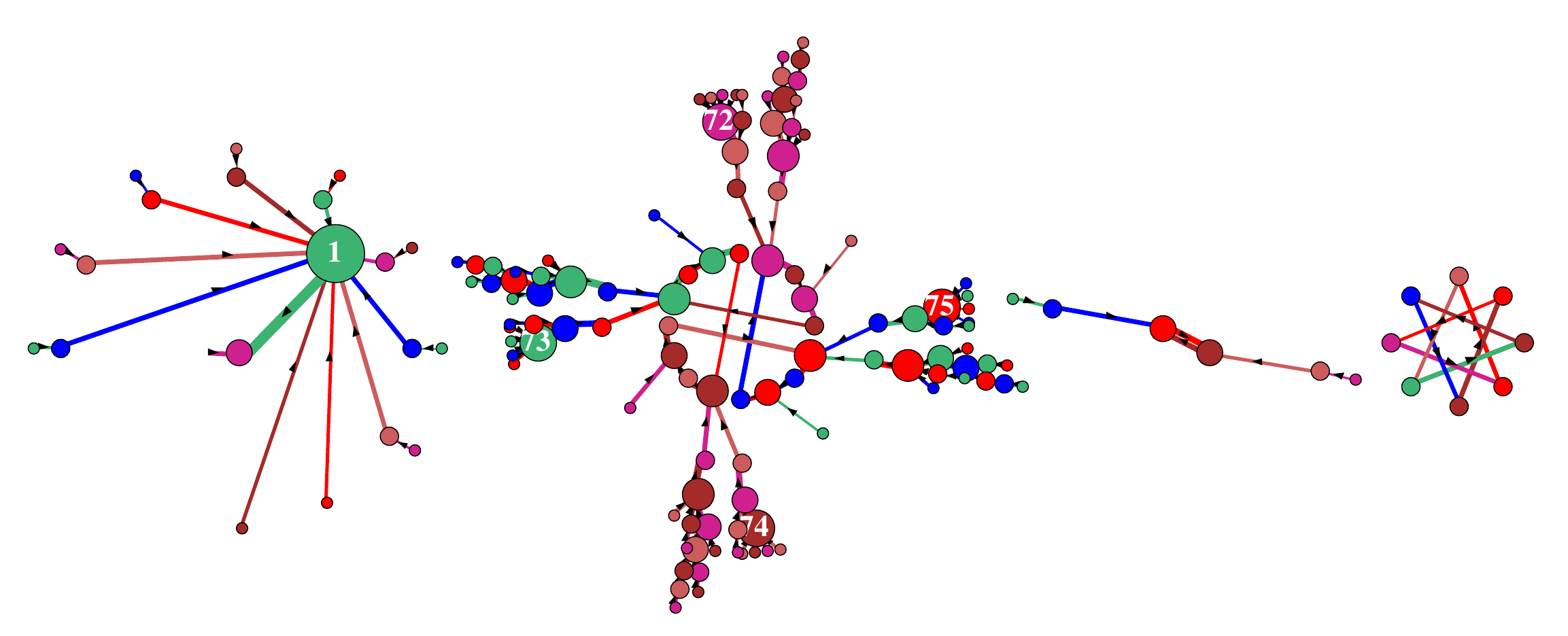}\\[-6ex]
   \begin{center}(b) istr-graph initially corresponds to the classic-graph layout,
   but here basins were dragged
   to avoid overlap. If compression reorders cycle states, cross links will be drawn,
   but are absent in the classic-graph.\end{center}
   \end{minipage}\\[.5ex]%layout r110istr.grh
   \begin{minipage}[t]{.85\linewidth}
     \includegraphics[width=1\linewidth, viewport=32 461 1163 776,clip]{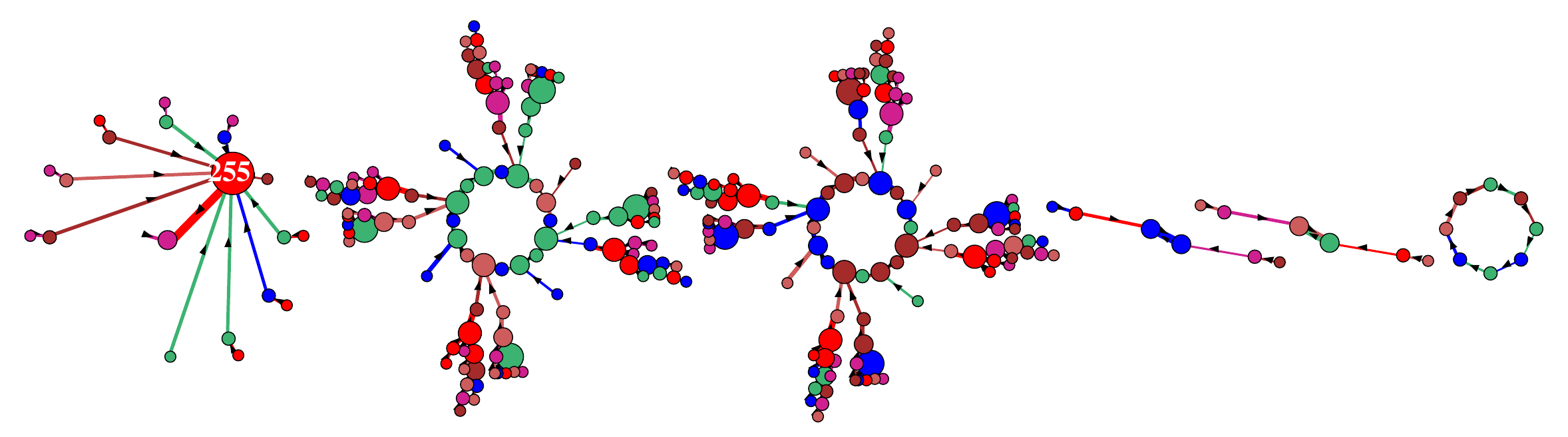}\\[-6ex]
   \begin{center}(c) ibaf-graph (similar to the uncompressed istr-graph) with $n^{8}$=256 nodes.
     Basins dragged to avoid overlap.\end{center}%layout r110ibaf.grh
    \end{minipage}
    }
    \end{center} 
\vspace{-2.5ex}
        \caption [The istr/ibaf graph CA $v2k3$ $n$=8 rcode 110]       
        {\textsf{Examples of the basin of attraction field, $v2k3$,
            $n$=8 1d CA rcode 110.  For the classic-graph (a) the
            layout\cite[\hspace{-1ex}\footnotesize{\#25}]{EDD} and
            compression\cite[\hspace{-1ex}\footnotesize{\#26.2}]{EDD}
            are preset, with the results inherited by the istr-graph
            (b).  The ibaf-graph (c) is very similar to the
            uncompressed istr-graph except for disc numbering.
            Initially nodes are shown as discs and both discs/links
            are scaled by inputs (in-degree). To illustrate the
            difference between disc numbering for istr/ibaf graphs,
            figure~\ref{uncompressed and compressed istr-graph}
            shows the 2nd basin in (a,b,c) limited to one backward level,
            and figure~\ref{r110istr-basin1-nframe.ps} shows the 1st
            basin in (b) an (c) with framed disc numbers.
            \label{r110ibaf.ps}}}
        
\end{figure}

\begin{figure}[t]
  \begin{center}
    \textsf{\small
      \begin{minipage}[t]{.85\linewidth}
  \includegraphics[width=1\linewidth,viewport=124 159 762 382,clip]{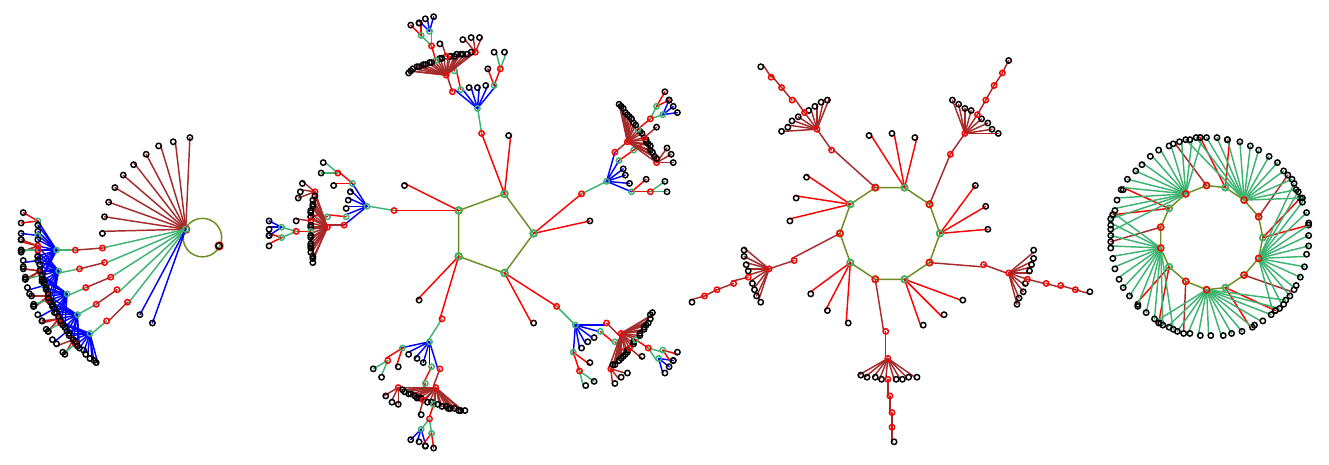}\\[-6ex]
  \begin{center} (a) Compressed basin of attraction field (classic-graph) of a binary 1d CA
    with 4 non-equivalent basins.\end{center}
  \end{minipage}\\[2ex]
  \begin{minipage}[t]{.85\linewidth}    
  \includegraphics[width=1\linewidth,viewport=155 290 1057 608,clip]{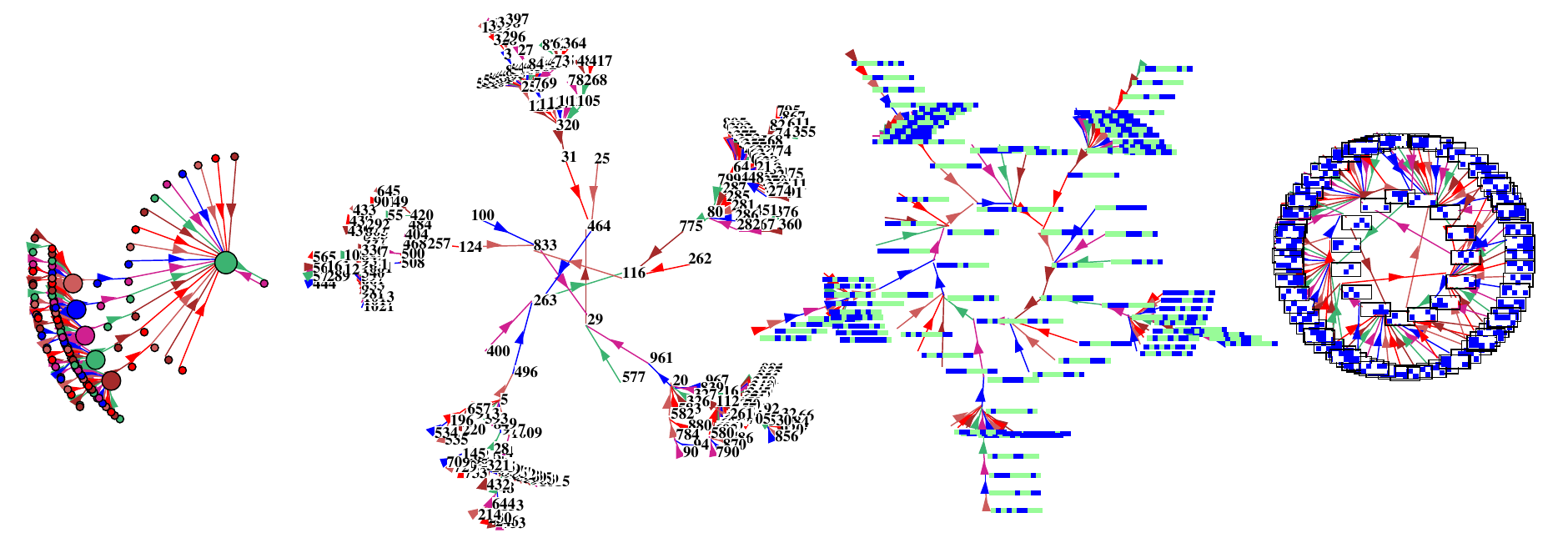}\\[-6ex]
  \begin{center} (b) istr-graph derived from (a) with each basin toggled to a different node display,\end{center}
  \end{minipage}\\[2ex]
  }
\end{center} %vector PostScript file
\vspace{-3ex}
\caption[A rearranged  istr/ibaf graphs]
        {\textsf{Compressed classic-graph (a) and  istr-graph (b) of the same
            binary 1d CA, $v2k3$, $n$=10, rcode=9. If uncompressed there would be
            7 basins, here reduced to the 4 non-equivalents.}}
\label{r9-ibaf}
\end{figure}

\mbox{Compression\cite{Wuensche92}\hspace{-.2ex}\cite[\hspace{-1ex}\footnotesize{\#26.2}]{EDD}}
(see also section~\ref{istr-graph compression of 1d CA dynamics}),
applicable for 1d CA with periodic boundaries, is a
shortcut in drawing single basins and the basin of attraction field more efficiently,
taking into account that any rotation
(within the bounds of rotation symmetry\footnote{ 
Rotation symmetry $s$ measures the maximum number of repeating segments
   into which the circular array can be divided,
   For an array size $n$ and segment size $g$, $s$=$n$/$g$.
   If $g$=$n$ (a disordered state, always the case for prime~$n$) then $s$=1.
   If  $g$=1 (a uniform state like 1111..) then $s$=$n$ its maximum value.
   If $s$$>$1 and $s$$<$$n$ the state is segmented with $g$$\leq$($n/2$). 
   It was shown in \cite{Wuensche92} that $s$ must be constant on the attractor,
   and may only increase on a transient.}
of the 1d CA circular array must be embedded in equivalent dynamics,
so rotation equivalent trees or
basins are computed from scratch by a reverse algorithm
only once, otherwise states are simply rotated within the
repeating segment of rotation symmetry. Just one of each prototype
basin is then drawn to represent the field,
and within each basin just one  prototype tree, or subtree for a
uniform state, is computed from scratch
(figures~\ref{r110ibaf.ps}ab, \ref{r9-ibaf});
computation is faster and the compressed graph conveys the essential information. 

\enlargethispage{4ex}
A directed graph of the field, single basin, or of just a subtree
rooted on a state, is drawn according to predefined graphic
\mbox{conventions\cite{Wuensche92}\hspace{-.2ex}\cite[\hspace{-1ex}\footnotesize{\#24-\#26}]{EDD}}.
These graphs are named ``classic'' as a distinction from
``interactive'' defined below.  Figures \ref{fig:rbn_P},
\ref{r110ibaf.ps}, \ref{r9-ibaf}, \ref{fig:quick_single_basin.ps},
\ref{fig:quick_subtree} show juxtaposed~examples.
\clearpage

\begin{figure}[b]
  %\vspace*{-2ex}
  \begin{center}
\begin{minipage}[t]{.85\linewidth}
\includegraphics[width=.89\linewidth,viewport=277 277 1181 579, clip]{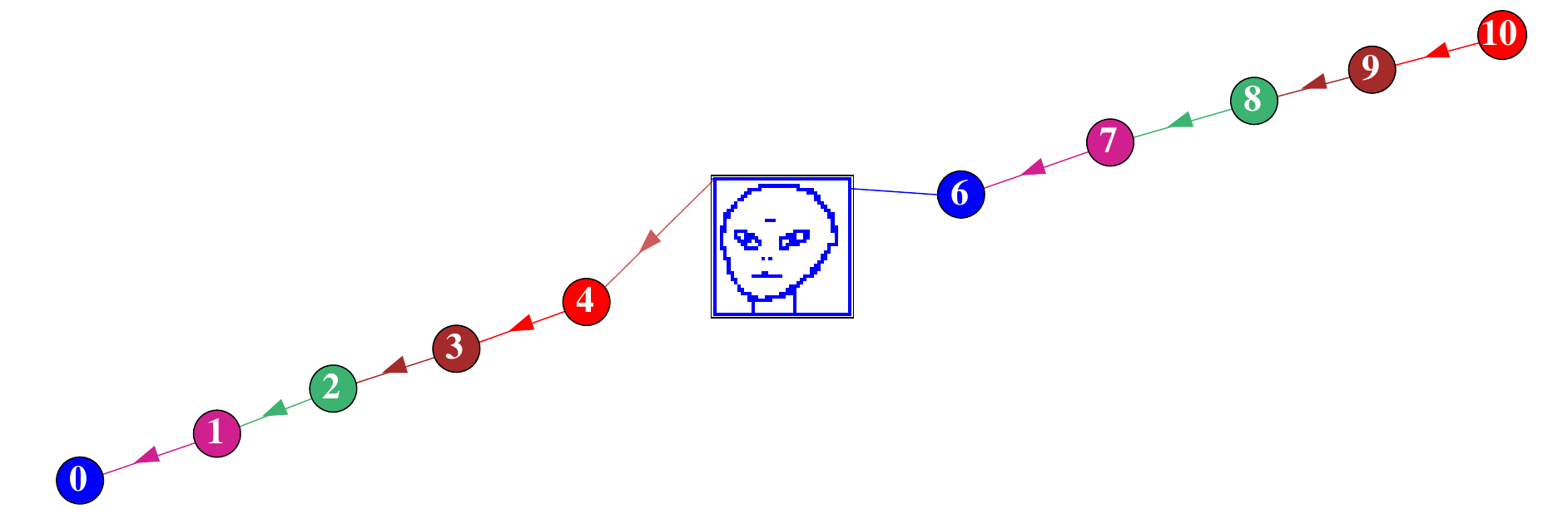}\\[-22ex]
\includegraphics[width=1\linewidth,viewport=187 218 1207 624, clip]{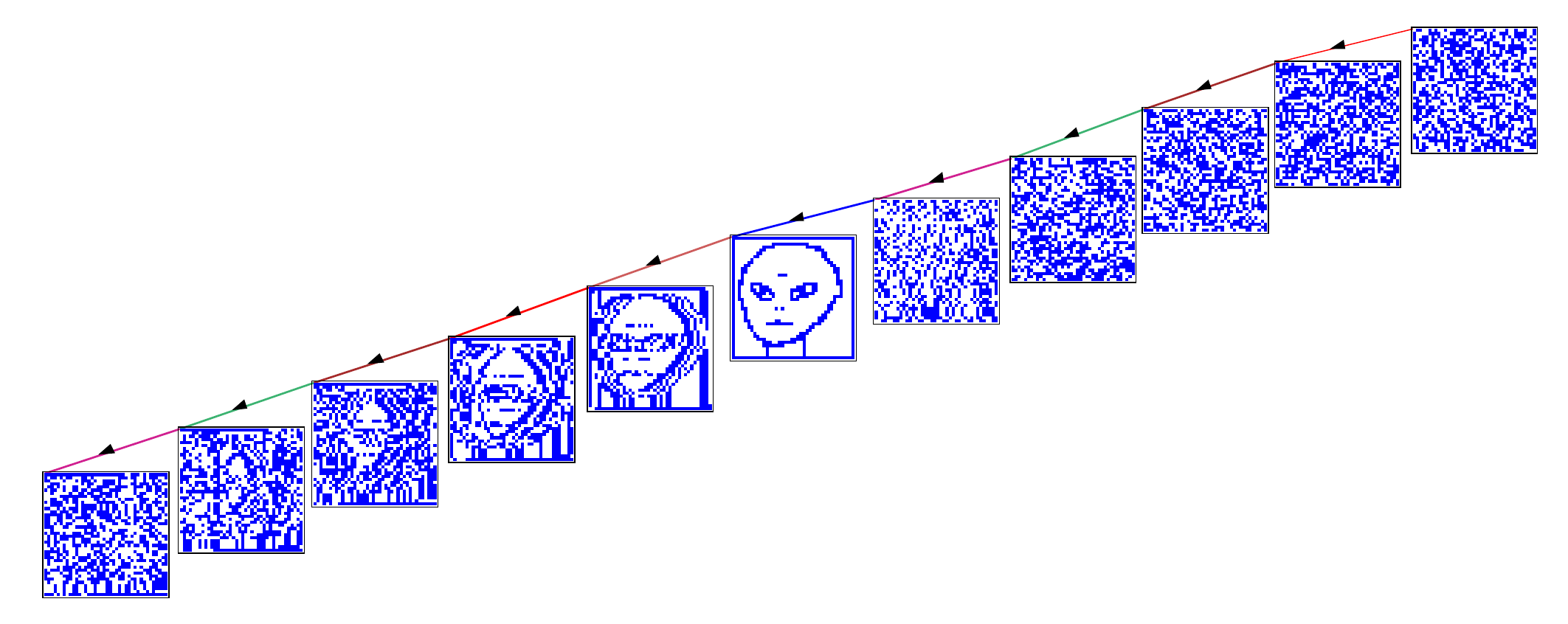}\\[-15ex]
\includegraphics[width=1\linewidth,viewport=18 223 1253 767, clip]{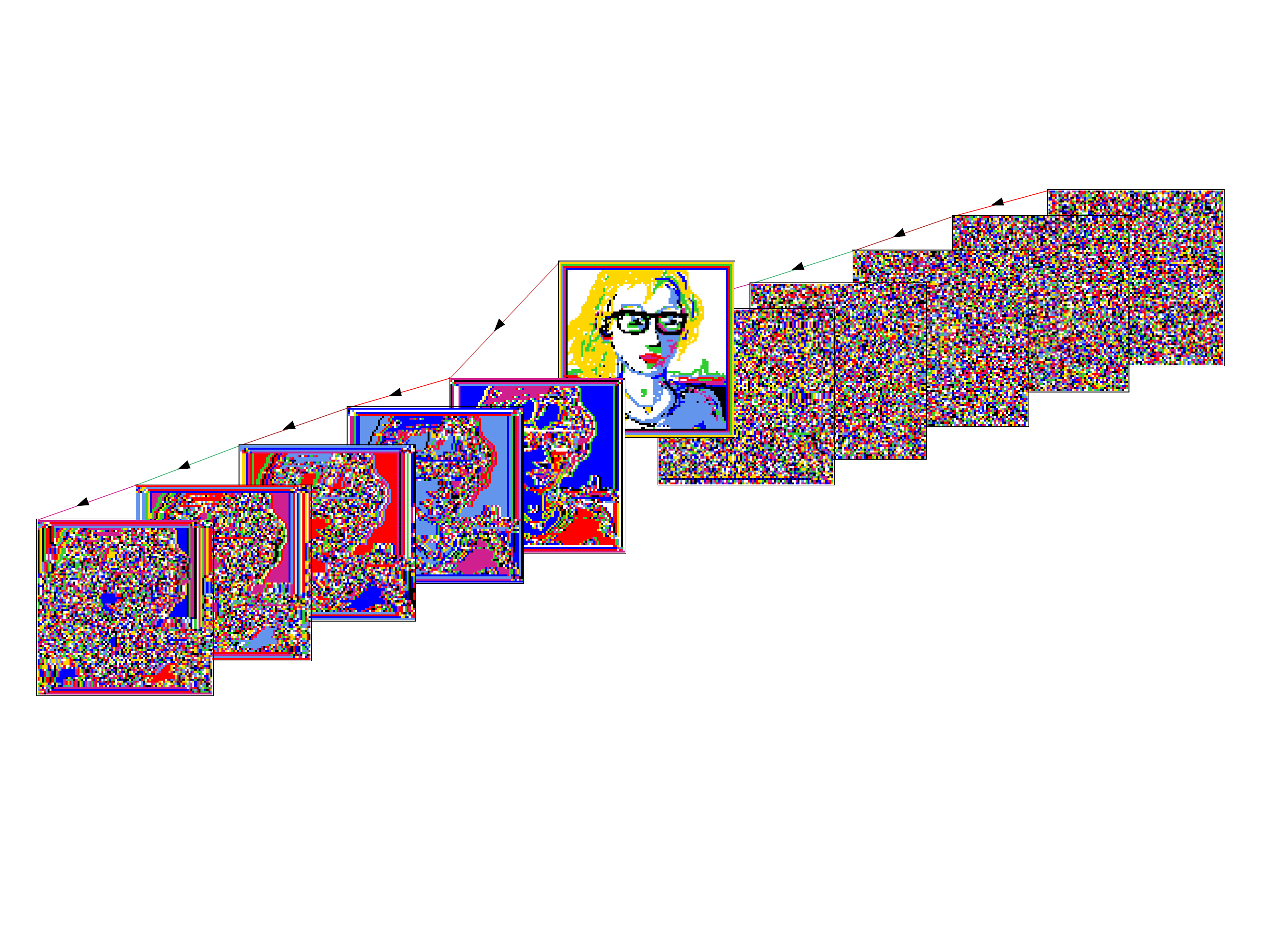}
\end{minipage}
\end{center}
\vspace{-3ex}
\caption[istr-graph of information hiding in chaos]
{\textsf{Information hiding in chaos within trajectories
of 1d CA chain-rules\cite{wuensche09} shown as istr-graphs, and
demonstrating how the istr-graph subtree
can deal with very large systems because the critical limiting metric is
the total number of nodes where disc numbers follow
the order of node computation, not the system state size ($2^{1600}$ and $8^{7744}$) 
which would make decimal numbering impossible.
Starting with a seed (centre) a classic subtree was
created\cite[\hspace{-1ex}\footnotesize{\#29.1-\#29.5}]{EDD}
by selecting
{\bf ``backward for subtree-b''}, then
{\bf ``forward before backwards?''} select 5 steps,
then {\bf ``limit backsteps-b''},
then \mbox{{\bf ``enter max backsteps ...:''}} enter 11.
Once drawn, the istr-graphs were adjusted with drag options.
\underline{\it Top}: $v$2 $k$7 40$\times$40 $n$=1600
2d ``Alien'', with numbered discs on either side.
\underline{\it Below Top}: the full ``Alien'' subtree in 2d.
\underline{\it Bottom}: $v$8 $k$4  88$\times$88 $n$=7744 ``Portrait''.
(\cite[\hspace{-1ex}\footnotesize{\#16.11}]{EDD} for classic-graph of these subtrees).
%figures~\ref{alien4-grey.ps}, \ref{face88f.ps}).
Arrows show the forward direction of time. 
\label{alien40o.ps}}}
\end{figure}

\clearpage
Although the default layout and presentation of a classic-graph can
with some slight effort be amended and adjusted before or even during
the actual drawing phase, once complete the classic-graph is static
and probably not exactly as one might wish. An interactive version of
the classic-graph opens up new perspectives for visualisation,
deconstruction and analysis.
In DDLab, interactivity for a directed graph includes the following attributes:
that a target node and its linked ``fragment'' defined by
inputs, outputs, or links in either direction, and a link distance,
can be relabelled, dilated, isolated, and dragged/dropped with elastic
or snap links. This has been the case for the well
established network and jump graphs\cite{DDLab2002update}, and for the
ibaf-graph\cite{wuensche2024} introduced recently.

Now the same interactivity is extended to the ``interactive state
transition graph'' (istr-graph) with an initial layout that
corresponds to its classic-graph precursor which can be
a subtree, single basin, or the whole field
(figures~\ref{alien40o.ps}, \ref{fig:rbn_P}, \ref{r9-ibaf}) 
and complies with the full range of classic-graph presentations and options,
such as compression, limiting
backward steps, forwards-before-backwards, pauses and interrupts. All
interactive functions for drag/drop, visualising, manipulating, deconstructing
and analysing apply as for the existing network/jump/ibaf graphs, with
some refinements made along with this istr-graph update.

The aspiration to manipulate the layout of the classic-graph provided
the motivation for the first successful implementation of its drag/drop
interactivity: the ``interactive basin of attraction field graph''
(ibaf-graph)\cite{wuensche2024}, adopting the interactive functions
and code for the network and jump graphs\cite{DDLab2002update}.  The
ibaf-graph (figure~\ref{r110ibaf.ps}c) is redrawn from a classic-graph computed exclusively by the
exhaustive reverse algorithm which holds the list of successors
(outputs) of every state in state-space, and draws the entire
uncompressed basin of attraction field with nodes numbered according
to their decimal states. Although the resulting ibaf-graph limits network
size and is less versatile than the istr-graph, it is nevertheless
retained for some unique attributes including a meaningful adjacency
matrix, and a geometric layout method in conjunction with the
jump-graph, and also as a reality check.

%.  Note that separate from the reverse
%algorithm there are two ``Classic-to-Interactive'' (C-to-I).
%algorithms.  The ibaf C-to-I is simpler than the istr-graph
%C-to-I but 

\begin{figure}[b] 
  \begin{center}
     \begin{minipage}[t]{.95\linewidth} 
     \textsf{\small
        \begin{minipage}[t]{.32\linewidth} %seed 152 r159 n8
           \includegraphics[width=1\linewidth,viewport=495 310 813 632,clip]{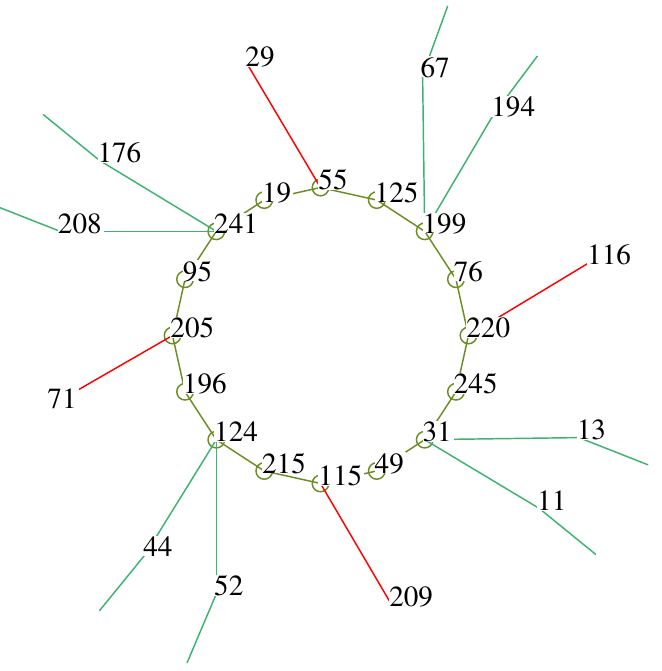}\\[-7ex]
             \begin{center} (a) uncompressed classic-graph in decimal\end{center}
        \end{minipage}
         \begin{minipage}[t]{.32\linewidth}
           \includegraphics[width=1\linewidth,viewport=495 310 813 632,clip]{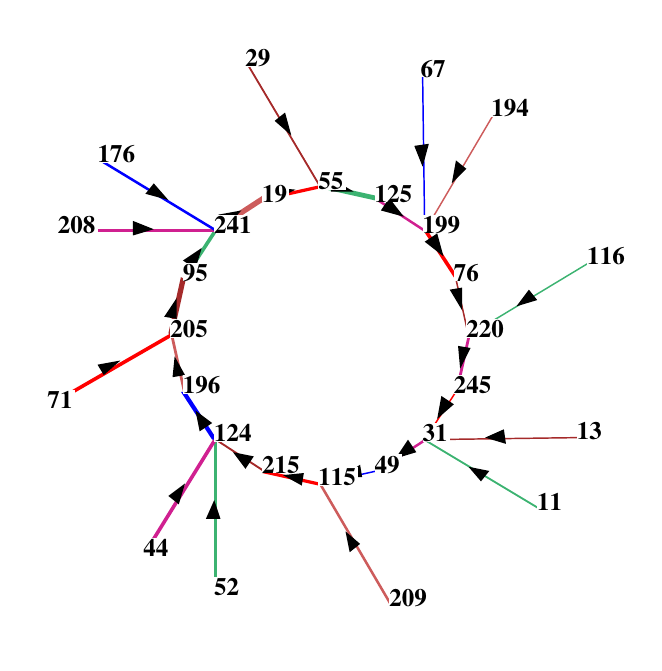}\\[-7ex]
             \begin{center} (b) uncompressed istr-graph in decimal\end{center}
        \end{minipage}
             \hfill
        \begin{minipage}[t]{.32\linewidth}
          \includegraphics[width=.95\linewidth,viewport=393 190 885 720,clip]{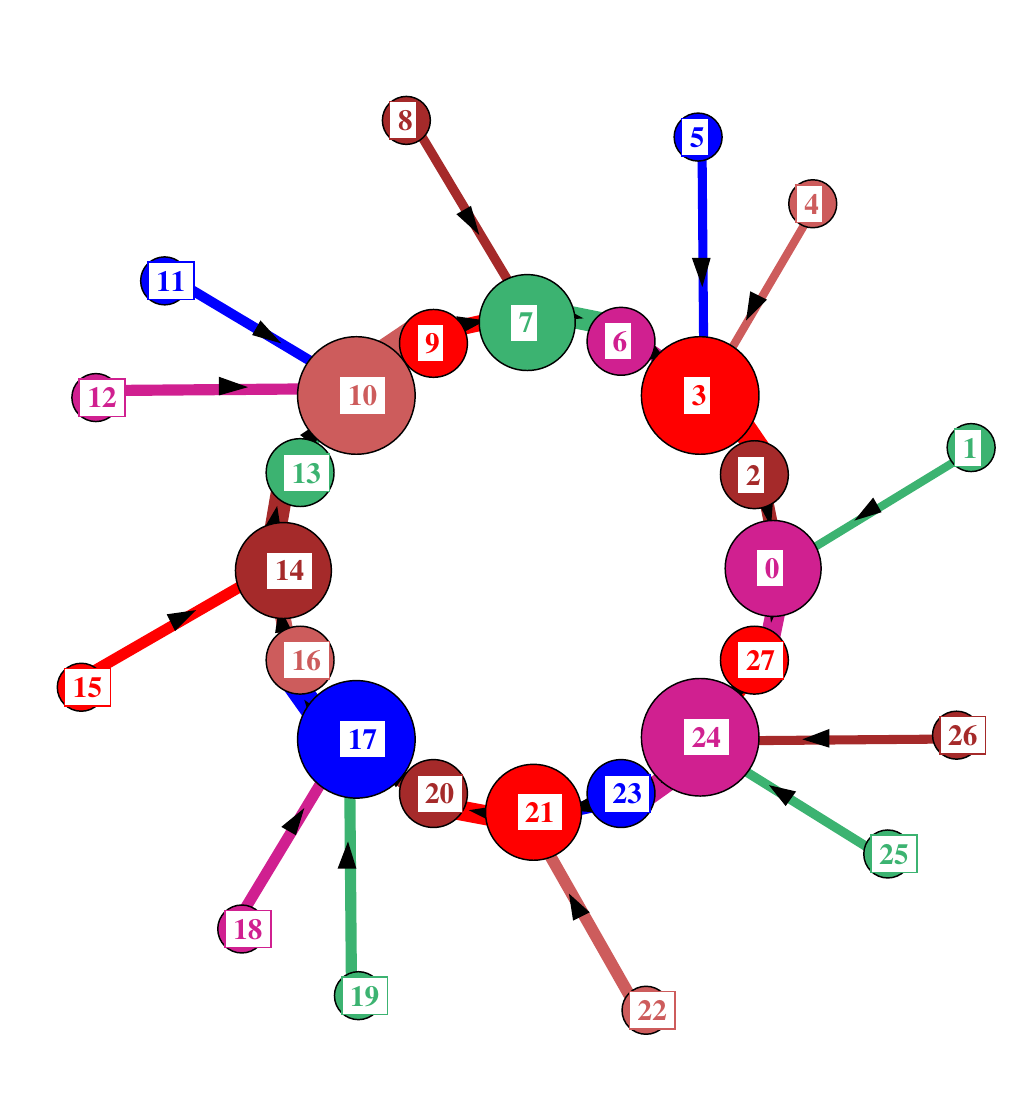}\\[-7ex]
             \begin{center} (c) uncompressed istr-graph as discs in computation order \end{center}
        \end{minipage}\\
        \begin{minipage}[t]{.32\linewidth}
          \includegraphics[width=1\linewidth,viewport=495 310 813 632, clip]{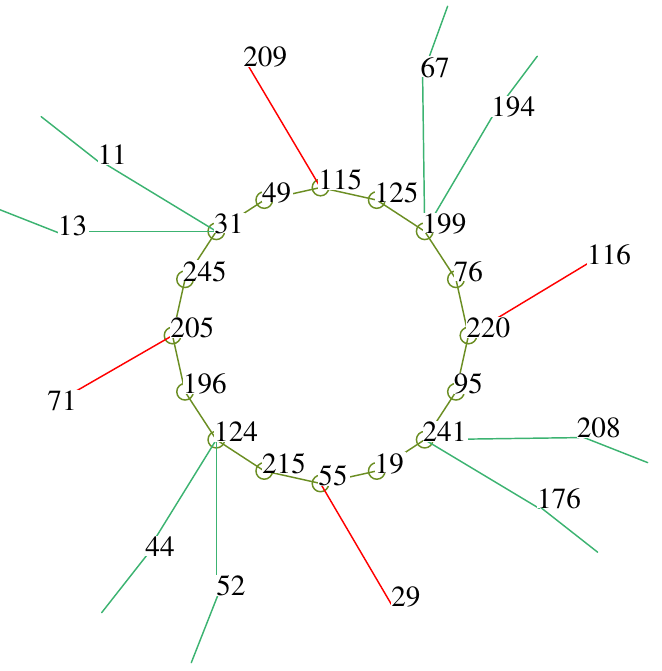}\\[-7ex]
             \begin{center} (d) compressed classic-graph in decimal, like (e) but no cross links\end{center}
        \end{minipage}
              \hfill
         \begin{minipage}[t]{.32\linewidth}
           \includegraphics[width=1\linewidth,viewport=495 310 813 632, clip]{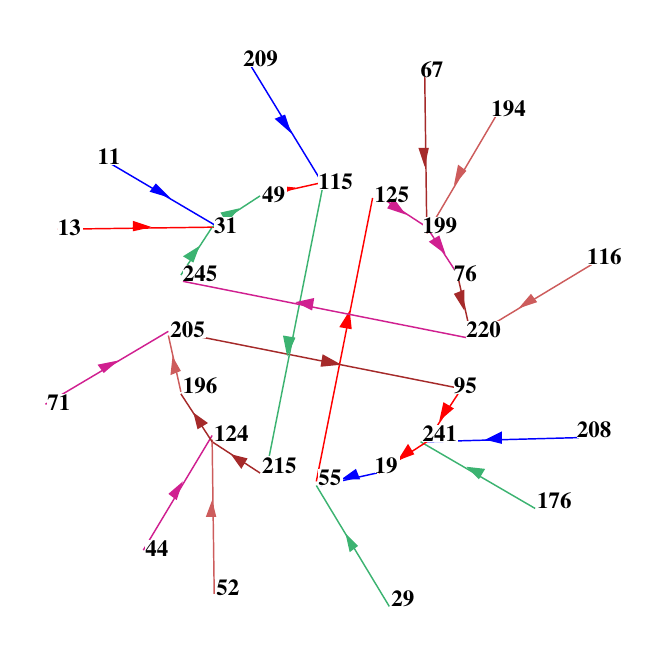}\\[-7ex]
             \begin{center} (e) compressed istr-graph in decimal includes cross links\end{center}
        \end{minipage}
              \hfill
         \begin{minipage}[t]{.32\linewidth}
           \includegraphics[width=.95\linewidth,viewport=393 220 885 720, clip]{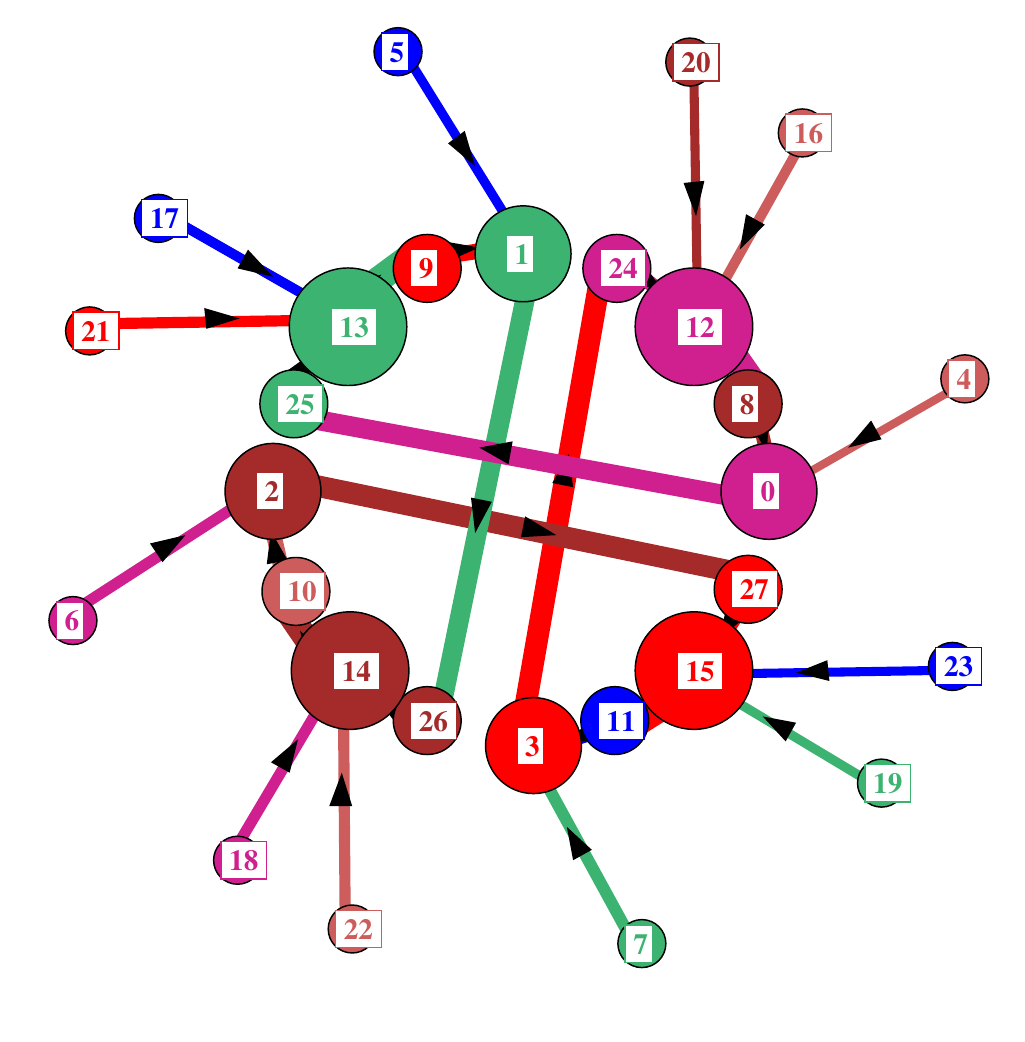}\\[-7ex]
             \begin{center} (f) compressed istr-graph as discs in computation order\end{center}
        \end{minipage}
     }
\end{minipage}     
\end{center}
        \vspace{-1ex} %\index{compression of CA dynamics!cross links}  \index{attractor cycle!cross links}  
       \caption[uncompressed and compressed classic and istr-graphs]
       {\textsf{Different presentations of the same single basin as a classic-graph
       and an istr-graph with backward time-steps limited to one level, the 2nd
       basin in figure~\ref{r110ibaf.ps}.
       \underline{\it Top row}: uncompressed,
       \underline{\it Bottom row}: compressed.
       States in (a,b,d,e) are shown in decimal, the same as ibaf-graph disc numbers.
       (c,f) show istr-graph disc numbers following the order of computation,
       here toggled with ``{\bf N}'' to put each number in a frame.
       In (d,e,f) compression has reordered nodes on the attractor cycle
       which is taken into account in the istr-graphs (e,f) to show the correct cross-links,
       whereas the compressed classic-graph (d) omits  cross links
       showing the attractor as a simple polygon. (1d CA $v2k3$, $n$=8 rcode 110 seed=245)
               \label{uncompressed and compressed istr-graph}}} 
\end{figure}

The new istr-graph is free of ibaf-graph state-space limitations
because its list of each graph state and its successor, required for
interactivity, is recorded one by one as the classic-graph is
drawn. Nodes are numbered by this order of computation (zero onward),
not as decimal states, though the corresponding state will appear if
toggled. This sequential method makes the flexibility of the
istr-graph possible for subtrees (figure~\ref{alien40o.ps}), single basins, compression,
interrupts, pauses, and so on, where the number of nodes is usually far
less than the size of \mbox{state-space}. The node number appears on
the ``disc'' version of the node display if the disk is big enough to
contain the number, but given that ``disc numbers'' are now more
significant for the istr-graph, an extra toggle has been added to show
framed disc numbers irrespective of disc size, as in
figure~\ref{r110istr-basin1-nframe.ps}.

For the istr-graph, any reverse
\mbox{algorithm\cite{Wuensche92}\hspace{-.2ex}\cite[\hspace{-1ex}\footnotesize{\#2.19}]{EDD}}
for pre-images of the three alternatives can be used; all allow classic-graph and
istr-graph compression. There are two direct reverse
algorithms: for 1d CA, or for systems with non-local wiring such as RBN.
For more general discrete dynamical networks (DDN) such as sequential updating and
random maps, the exhaustive algorithm is available (required for the
ibaf-graph). This allows a reality check where the RBN reverse
algorithm can compute 1d CA, and the DDN reverse algorithm can compute
both CA and RBN. Likewise, the istr-graph and ibaf-graph based on
the same classic-graph must give the same results, though with
different disc numbers as mentioned above, but with the same decimal
states when toggled.

With the aid of example figures, this article introduces the new
istr-graph.
Section \ref{A brief mention of the network-graph and jump-graph} briefly
mentions the network and jump graphs.  
Section \ref{The ist-graph and ibaf-graph} describes the
scope, comparison, and activation of the istr-graph and
ibaf-graph.
Sections \ref{Interactive graph reminders}  onward
describe the various interactive functions, most of which are common
to all four interactive graphs: network, jump, ibaf, and istr.

%---------------------------------------------------------------------

\section{A brief mention of the network-graph and jump-graph}
\label{A brief mention of the network-graph and jump-graph}

\noindent {The network-graph and jump-graph
  have been interactive features in DDLab since the 2002\cite{DDLab2002update}
  and are further documented in \cite{wuensche2024}\hspace{-.2ex}\cite[\hspace{-1ex}\footnotesize{\#20}]{EDD}.
  A very brief summery follows.

  The network-graph\cite[\hspace{-1ex}\footnotesize{\#20.5}]{EDD}
  represents the CA, RBN or DDN network itself,
and can be initiated at various stages when the underlying network is
defined, for example at the "wiring
graphic"\mbox{\cite[\hspace{-1ex}\footnotesize{\#17.4}]{EDD}}.  The
weight of discs/links can toggle between inputs or outputs.  Links can
be omitted to run space-time patterns within a network-graph layout,
and links can be added and cut to create an arbitrary network
presentation but without affecting the underlying
network\footnote{Adding/cutting links in any interactive graph is confined within the graph
and does not affect the underlying setup outside the graph context.}.

The jump-graph\cite[\hspace{-1ex}\footnotesize{\#20.7}]{wuensche2002,Voorhees,EDD}\footnote{The jump-graph was originally named the ``meta-graph''\cite{DDLab2002update}.} represents the
probability of jumping between basins of attraction due to 1-bit (or 1-value)
perturbations to attractor states. Nodes shown as discs
are scaled by basin volume and edges by jump probability.
There are two types, the
\mbox{{\it f}-jump-graph\cite[\hspace{-1ex}\footnotesize{\#31.7.1}]{EDD}}
directly from the classic uncompressed basin of attraction field,
or for larger systems the \mbox{{\it h}-jump-graph} \cite[\hspace{-1ex}\footnotesize{\#31.7.2}]{EDD}}
from the ``attractor histogram''\cite[\hspace{-1ex}\footnotesize{\#31.7.8}]{EDD}
derived statistically from space-time patterns.
Classic basins can be drawn at {{\it f}-jump-graph nodes, and the resulting layout saved,
then loaded back into an ibaf-graph, providing geometric or arbitrary layout
possibilities\cite[\hspace{-1ex}\footnotesize{\#20.14}]{EDD}.

%---------------------------------------------------------------------
\section{The istr-graph and ibaf-graph}
\label{The ist-graph and ibaf-graph}

\noindent The new istr-graph introduced in this article
redraws {\it any}\footnote{A limiting factor is
the number of nodes in the classic-graph which may be more than the
istr-graph can handle.} ``classic'' attractor basin
(classic-graph) created in DDLab as an interactive
image, where ``istr'' stands for
``\underline{\bf i}nteractive \underline{\bf s}tate \underline{\bf tr}ansition''.
This is an important and necessary update on the pre-existing
``\underline{\bf i}nteractive \underline{\bf b}asin of
\underline{\bf a}ttraction \underline{\bf f}ield''
\mbox{ibaf-graph}\cite{DDLab2023update,wuensche2024}
which is limited to just the uncompressed field.

The two graph types,
contrasted in section~\ref{Contrasting the istr-graph versus ibaf-graph},
are directed graphs with out-degree=1 and variable
in-degree$\geq$0 --- pre-images --- which sets the weight of
discs/links, and where disc numbers for dragging nodes follow
computation order for the istr-graph, independent of state-space
\mbox{(figures~\ref{alien40o.ps}, \ref{node-dis.ps})}, in contrast to the
ibaf-graph disc numbers that reflect the entire state-space
(figure~\ref{r110ibaf.ps}c).  Both graphs are subject to interactive
functions for graph manipulation, deconstruction and analysis
described in sections \ref{Interactive graph reminders}---\ref{Drag-graph options}.
This section describes and compares the scope of the istr/ibaf graphs, and the
methods for activation.
\clearpage

\begin{figure}[htb]
%\vspace{-5ex}
   \begin{center}
   \includegraphics[width=.8\linewidth,viewport=144 13 1208 932,clip]{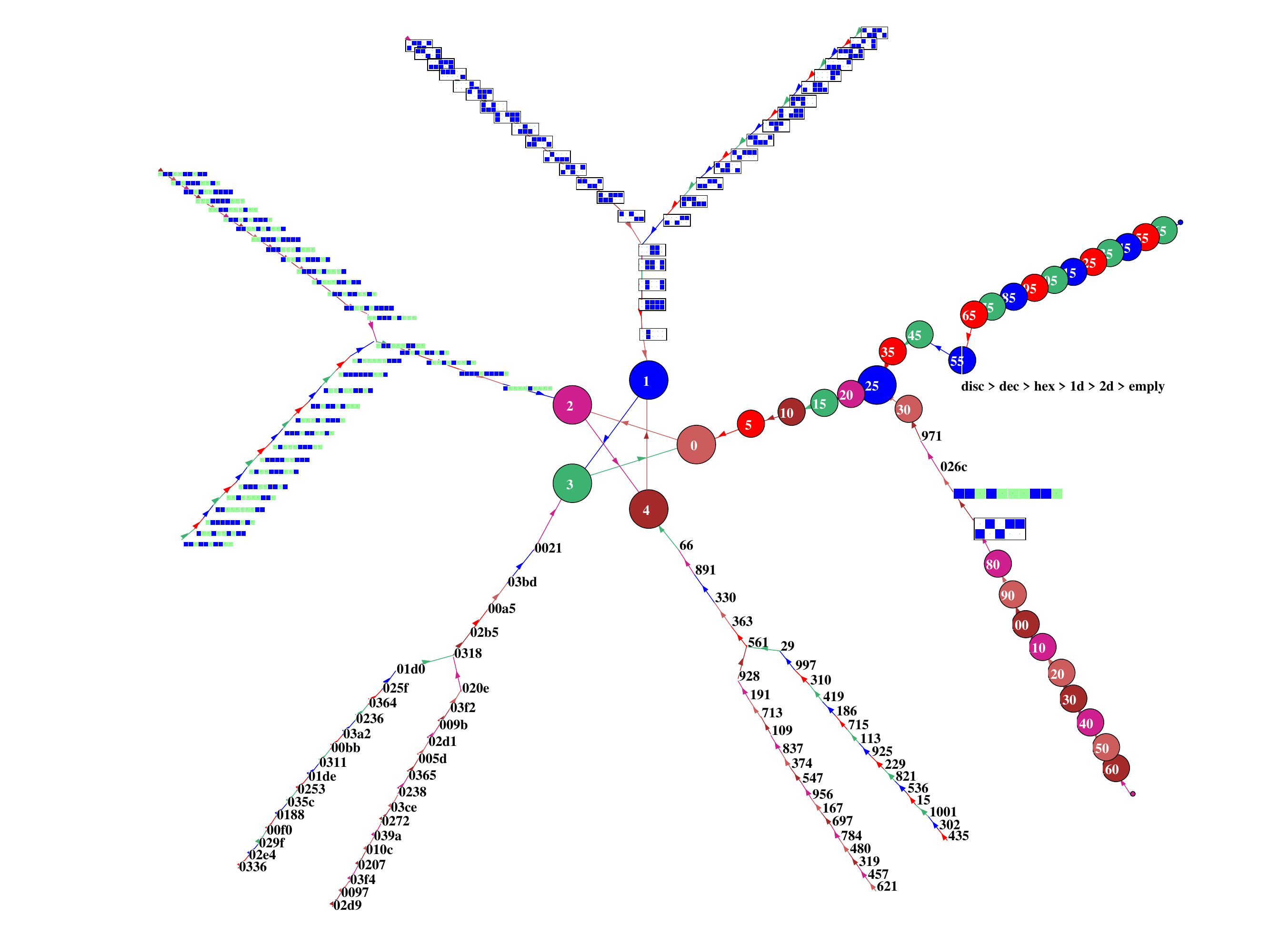}
   \end{center} 
    \vspace{-4ex}   
    \caption [Node displays --- istr/ibaf graphs]
    {\textsf{Illustrating different node displays for an istr-graph in drag mode,
    where``disc numbers'' follow the sequence of node computation not the decimal state.
    In the ibaf-graph, disc and decimal state numbers are the same.
    This example shows a single basin with compression of equivalent subtrees
    which may reorder attractor states, as in this case, accounting for cycle cross links.
    (1d CA $v2k3$ $n$=10 rcode 30). 
    \underline{\it Lower Right}: Examples of 5 alternative displays within
    a transient (``empty'' is skipped).
    \underline{\it Upper Right}: a created
    ``label''\mbox{\cite[\hspace{-1ex}\footnotesize{\#20.12}]{EDD}}
     adjacent to disc 55, noting the toggling sequence of the 6 alternatives node displays.
     In other transients nodes were toggled to dec, hex, and  1d/2d patterns.
     \label{node-dis.ps}}}    
\end{figure}

\begin{figure}[htb]
%\vspace*{-2ex}
\begin{center}
\small
\begin{minipage}[t]{.8\linewidth}
\begin{minipage}[t]{.45\linewidth}
  \includegraphics[height=1\linewidth,viewport=123 276 433 560,clip]{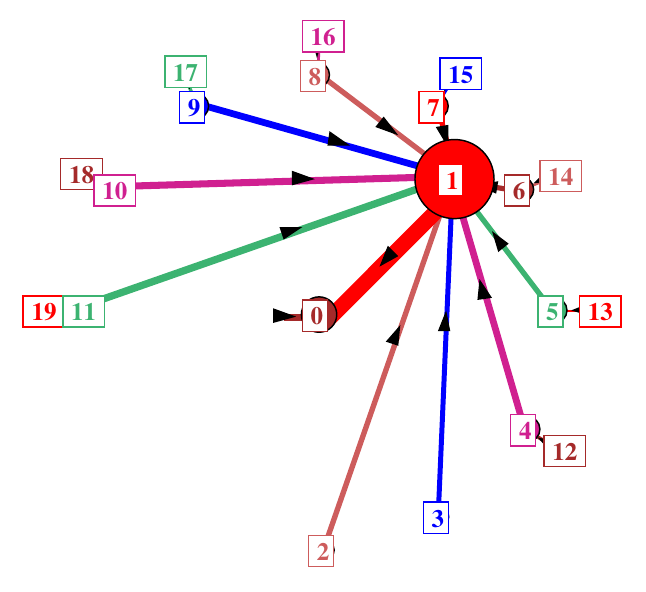}\\[-10ex]
\textsf{\phantom{-----}(a) istr-graph}\\
\textsf{\phantom{-----------}basin}
\end{minipage}
\hfill
\begin{minipage}[t]{.45\linewidth}
  \includegraphics[ height=1\linewidth, viewport=223 248 563 556,clip]{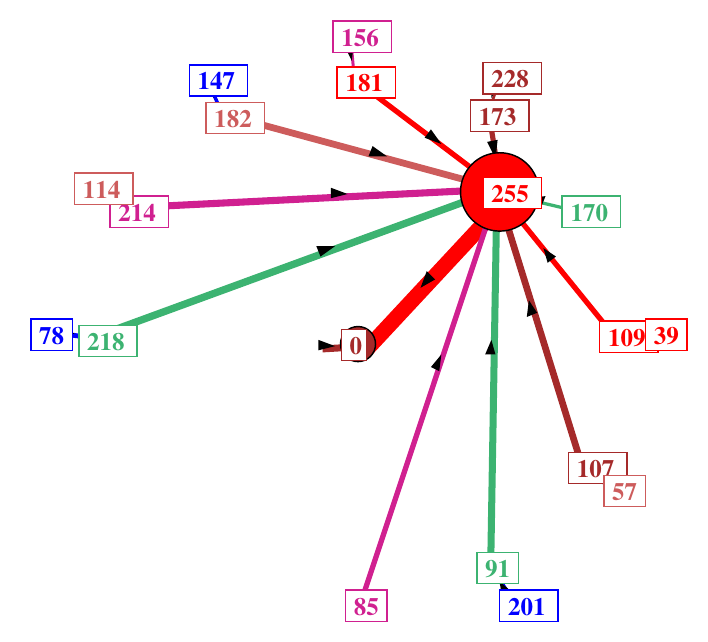}\\[-10ex]
\textsf{\phantom{-----}(b) ibaf-graph}\\
\textsf{\phantom{-----}isolated basin}
\end{minipage}
\end{minipage}
\end{center}  
\vspace{-4ex}
\normalsize
\caption[istr/ibaf disc numbers]
{\textsf{Disc numbers, which also feature in the drag-reminder, differ for istr/ibaf graphs.
The istr-graph disc numbers (a) follow the computation order (zero onward) of successive nodes
independent of the system state size which can be very large as in figure~\ref{alien40o.ps},
and where the critical limiting metric is the total number of nodes.
By contrast, the ibaf-graph disc numbers (b) are the same as the state decimal equivalents, 
To illustrate the difference, this example shows
the first basin in figure~\ref{r110ibaf.ps}(b,c).
The alternative framed disc number design was set with the toggle ``{\bf nodes-N}''
in the initial-graph.
\label{r110istr-basin1-nframe.ps} }}
\end{figure}

\subsection{Contrasting the istr-graph versus ibaf-graph}
\label{Contrasting the istr-graph versus ibaf-graph}

\noindent Any classic-graph computed and drawn by any reverse algorithm can be the
precursor of an istr-graph, including a subtree, single basin, or basin of
attraction field, allowing compression for \mbox{1d CA}, limits
to backward steps, ``forward before backward'', pauses, interrupts, and
other functions in \mbox{\cite[\hspace{-1ex}\footnotesize{\#23-\#30}]{EDD}}.

Because of this more flexible and varied repertoire, the istr-graph
largely supersedes the \mbox{ibaf-graph}, limited as it is to the uncompressed
basin of attraction field computed by the exhaustive algorithm.
Nevertheless the ibaf-graph includes some useful attributes not
available in the istr-graph, as follows,

\begin{enumerate}
\item can combine with the jump-graph to preset
   basins positions in arbitrary or geometric layouts, including circular,
   spiral, 1d/2d/3d\cite[\hspace{-1ex}\footnotesize{\#20.14.1}]{EDD}.\\[-4ex]
\item numbers nodes by state-space decimal values as in the classic-graph,
  a more natural numbering scheme for the basin of attraction field than
  by computation order for the istr-graph.\\[-4ex]
\item allows a meaningful adjacency
  matrix\mbox{\cite[\hspace{-1ex}\footnotesize{\#20.19.3}]{EDD}}
  because of state-space disc numbering above.
  For the istr-graph, matrix results are trivial because of its
  computation order numbering.
\end{enumerate}

In addition, the ibaf-graph is computed from the classic-graph with a
different and simpler algorithm to the istr-graph, so provides an important
reality check. For these reasons, and that it works well as conceived, the ibaf-graph
is retained in DDLab, but is limited vis-\`{a}-vis the istr-graph as follows,
%vis-\`{a}-vis the istr-graph as follows,

\begin{enumerate}[resume]
\item applies only to a complete and uncompressed basin of attraction field.\\[-4ex]
\item cannot deal with compression.\\[-4ex]
\item cannot deal with single basins or subtrees, though these can be
  isolated from the basin of attraction field\cite[\hspace{-1ex}\footnotesize{\#20.4.10}]{EDD}.\\[-4ex]
\item uses the exhaustive reverse algorithm with its system size
      limits \mbox{\cite[\hspace{-1ex}\footnotesize{\#29.7}]{EDD}}\footnote{For the ibaf-graph,
     these limits are much greater than the modest sizes required for practical purposes.}, and
     extra steps to activate.
\end{enumerate}

%\enlargethispage{\baselineskip} %equi to raggedbottom
\subsection{istr/ibaf graph disc numbers}
\label{istr/ibaf graph disc numbers}

\noindent Discs are numbered from 0 (zero) onward for both the istr/ibaf graphs,
but whereas ibaf-graph disc numbers are the same as a node's decimal state,
this is not the case for the istr-graph where disc numbers follow the
order of computation (figure~\ref{r110istr-basin1-nframe.ps}).
This is necessary because the istr-graph usually deals with a classic-graph with fewer
nodes than state-space, though the decimal state is
one of the node displays which can be toggled with
key ``{\bf =}'' (figures~\ref{node-dis.ps}, \ref{istr-graph single basin}).

The disc numbers feature in the drag-reminder to identify
the active node, to cut/restore links to the active node,
to cut/add/restore links between two nodes, and
in defining and dragging blocks.

\subsection{Scope of the istr-graph}
\label{Scope of the istr-graph}

Below is a summary of the scope of the istr-graph,
and its respect of its classic-graph precursor,

\begin{s_enumerate}
\setlength\itemsep{1ex}
\item applies to any type of ``attractor basin'' or classic-graph, including a subtree,
   single basin, or the basin of attraction field.
   Note that a classic-graph may have more nodes than the istr-graph can handle in terms of
   computation time, and subsequent dragging ability.
\item is computed from any classic-graph with an equally fast
   but more complex algorithm than the ibaf-graph. %explaneing why it postdated the ibaf-graph)  

 \item any of the three reverse algorithms for the classic-graph\cite[\hspace{-1ex}\footnotesize{\#2.19}]{EDD}
    is applicable, but a direct reverse algorithm makes activation simpler.

\item respects ``compression''\cite[\hspace{-1ex}\footnotesize{\#26.2}]{EDD}
   by equivalent trees and basins for
   1d\footnote{For 2d the istr-graph does not support compression, though the
  classic-graph does to some extent.} CA with periodic boundaries
  (where compression is the default), and will show attractor cross-links (if any)
  as in figure~\ref{uncompressed and compressed istr-graph}(e,f).

\item with compression set, respects suppressing copies of equivalent
  trees\cite[\hspace{-1ex}\footnotesize{\#26.2.3}]{EDD},
  and showing garden-of-Eden nodes within the suppressed copies (figure~\ref{r110-GofE-istr.ps}).

\item  respects running ``forward before backwards'' for subtrees\cite[\hspace{-1ex}\footnotesize{\#29.2}]{EDD}
   (figure~\ref{alien40o.ps}).

\item  respects a partly drawn classic-graph that is
   paused\mbox{\cite[\hspace{-1ex}\footnotesize{\#27.1.4}]{EDD}}
   (figure~\ref{basin-pause-istr.ps}) or
   interrupted\mbox{\cite[\hspace{-1ex}\footnotesize{\#30.2}]{EDD}}
    (figure~\ref{EaryExit110-istr.ps}).

 \item  respects limiting backward time-steps\cite[\hspace{-1ex}\footnotesize{\#29.5}]{EDD}
   for subtrees (figure~\ref{fig:quick_subtree}) and single basins
   (figure~\ref{uncompressed and compressed istr-graph}), and
   bare attractors if the limit is set to zero (figure~\ref{Gosper38x16-istr.ps}).
\end{s_enumerate}

\begin{figure}[t]
  \begin{center}
    \begin{minipage}[b]{.85\linewidth}
\begin{minipage}[b]{.62\linewidth}
\includegraphics[width=1\linewidth,viewport=210 107 865 669,clip]{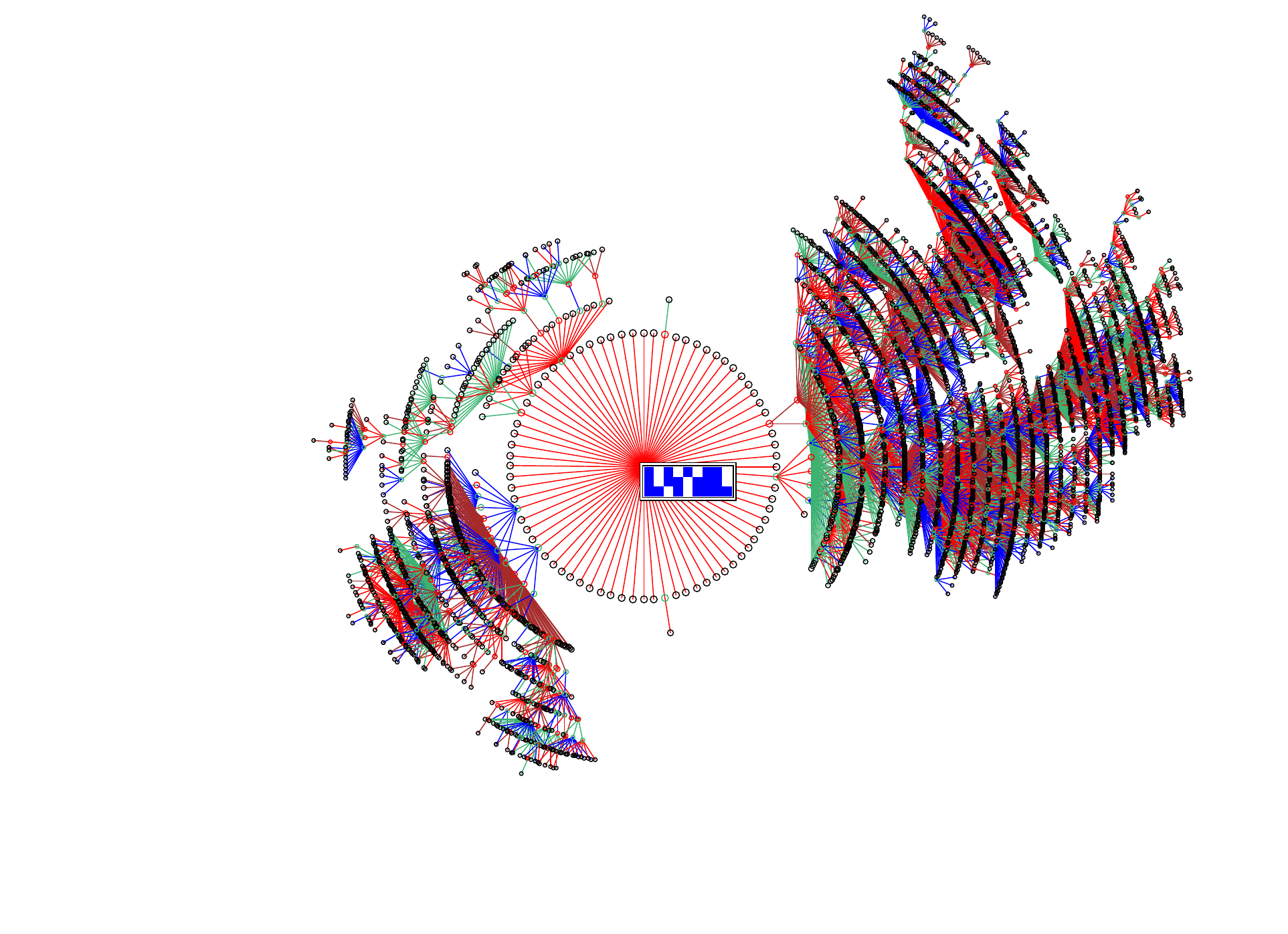}\\[-4ex]
\phantom{xxxxxxxxxxxxxxxxxxxx}\textsf{\small (a) classic-graph}
\end{minipage}
\hfill %.73
\begin{minipage}[b]{.33\linewidth}%{.33\linewidth}
\fbox{\includegraphics[width=1\linewidth,viewport=358 92 939 769,clip]{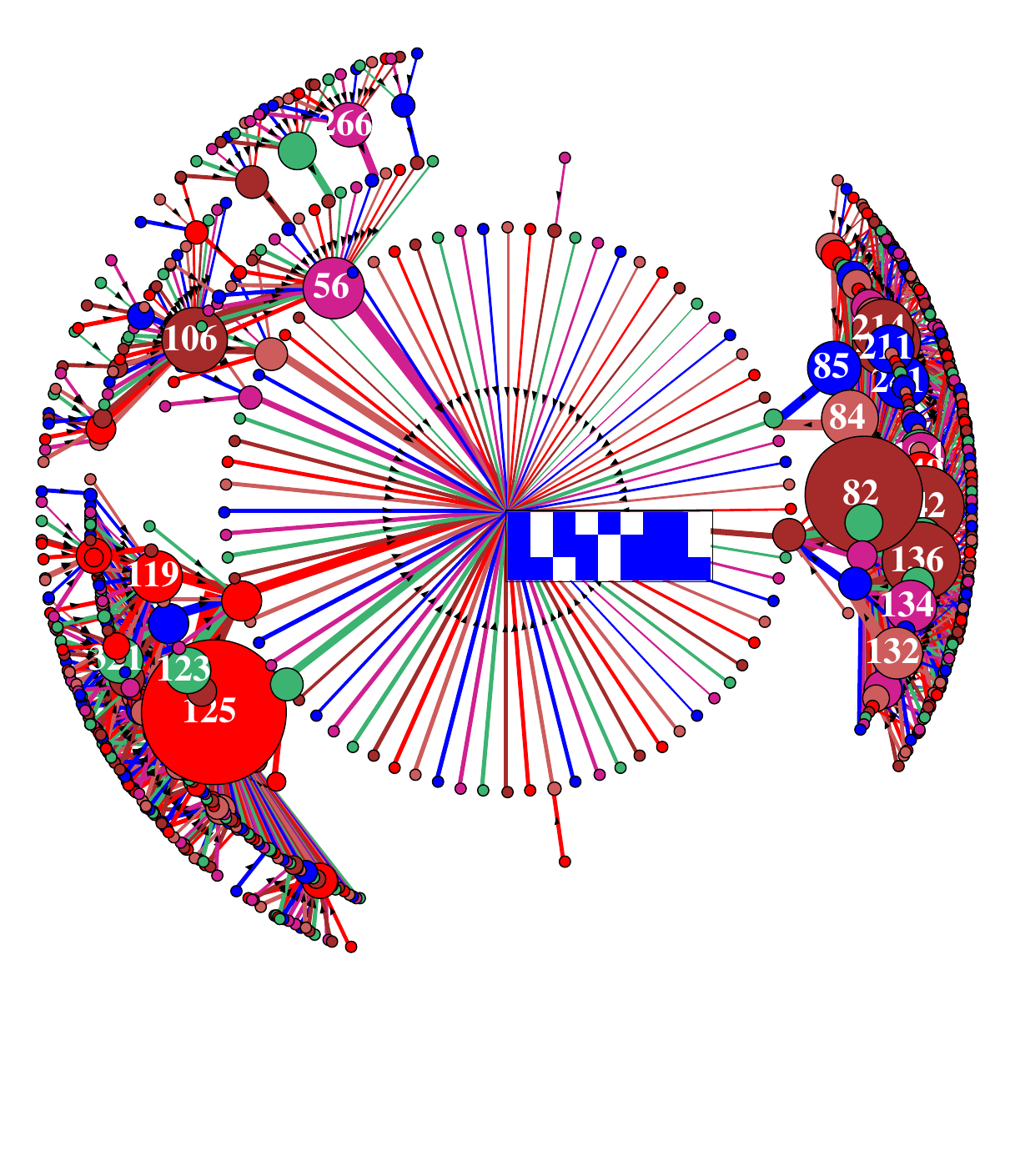}}\\[-4ex]
\phantom{x}\textsf{\small \mbox{(b) istr-graph, set to 5 levels}}
\end{minipage}
    \end{minipage}
\end{center}    
\vspace{-2ex}
\caption[Classic and istr-graph subtrees] 
        {\textsf{A subtree from
            the central seed shown as a 9$\times$3 bit pattern (hex 055addaf)..
            (a) the complete classic-graph with 11324 states/nodes.
            (b) the istr-graph limited to 5 backward levels, with 1050 states/nodes, respecting the
                limit set in the classic graph (1d CA, $v2k5$, $n$=27, hex rcode 5afbb1ae).
      \label{fig:quick_subtree}}}
  \vspace{-2ex}
\end{figure}

\begin{figure}[b]
  \begin{center}   
   \begin{minipage}[b]{.9\linewidth}
       \begin{minipage}[b]{1\linewidth}
         \includegraphics[width=.48\linewidth,viewport=282 115 994 835,clip]{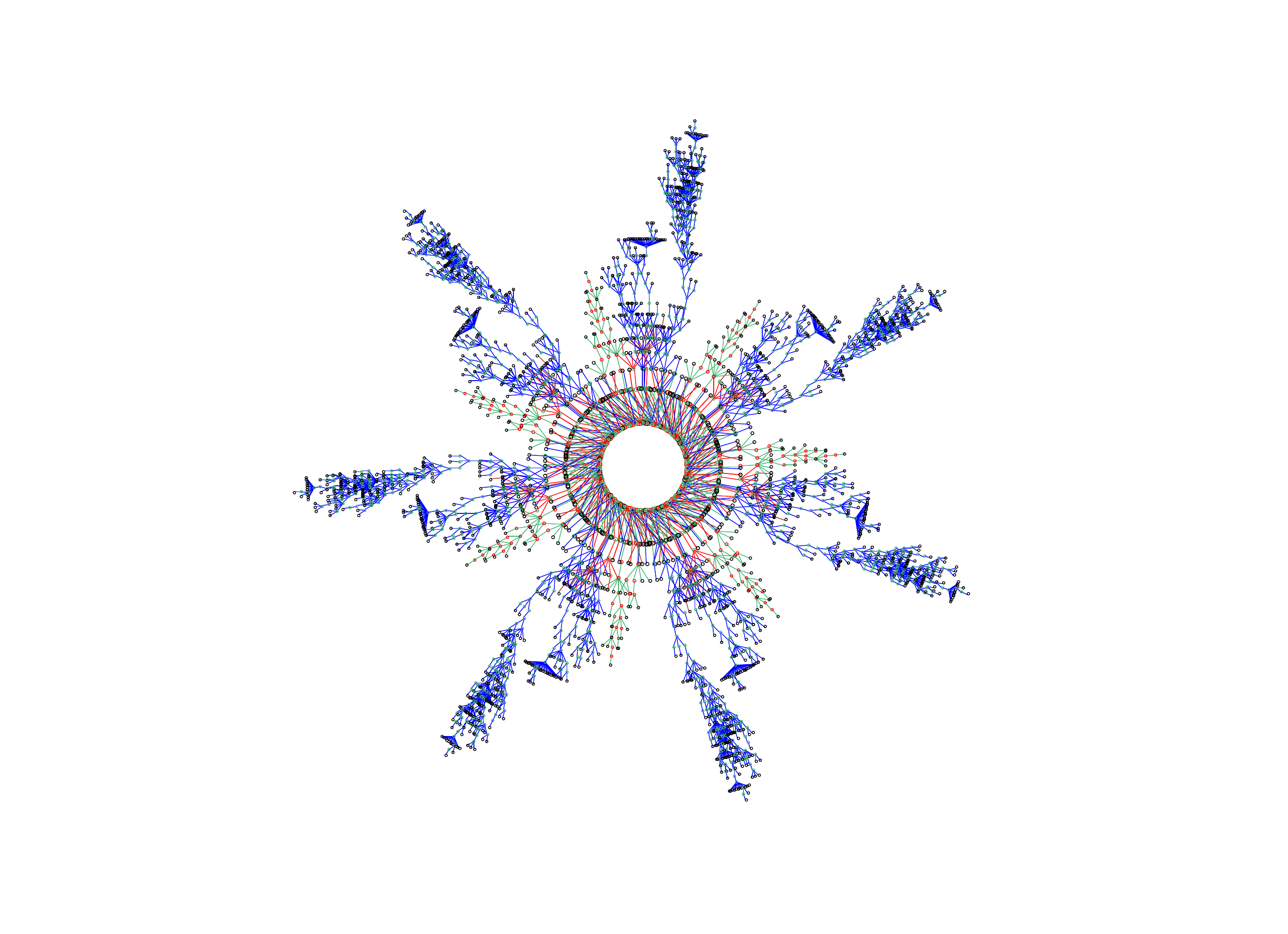}
      \hfill
      \includegraphics[width=.48\linewidth,viewport=282 115 994 835,clip]{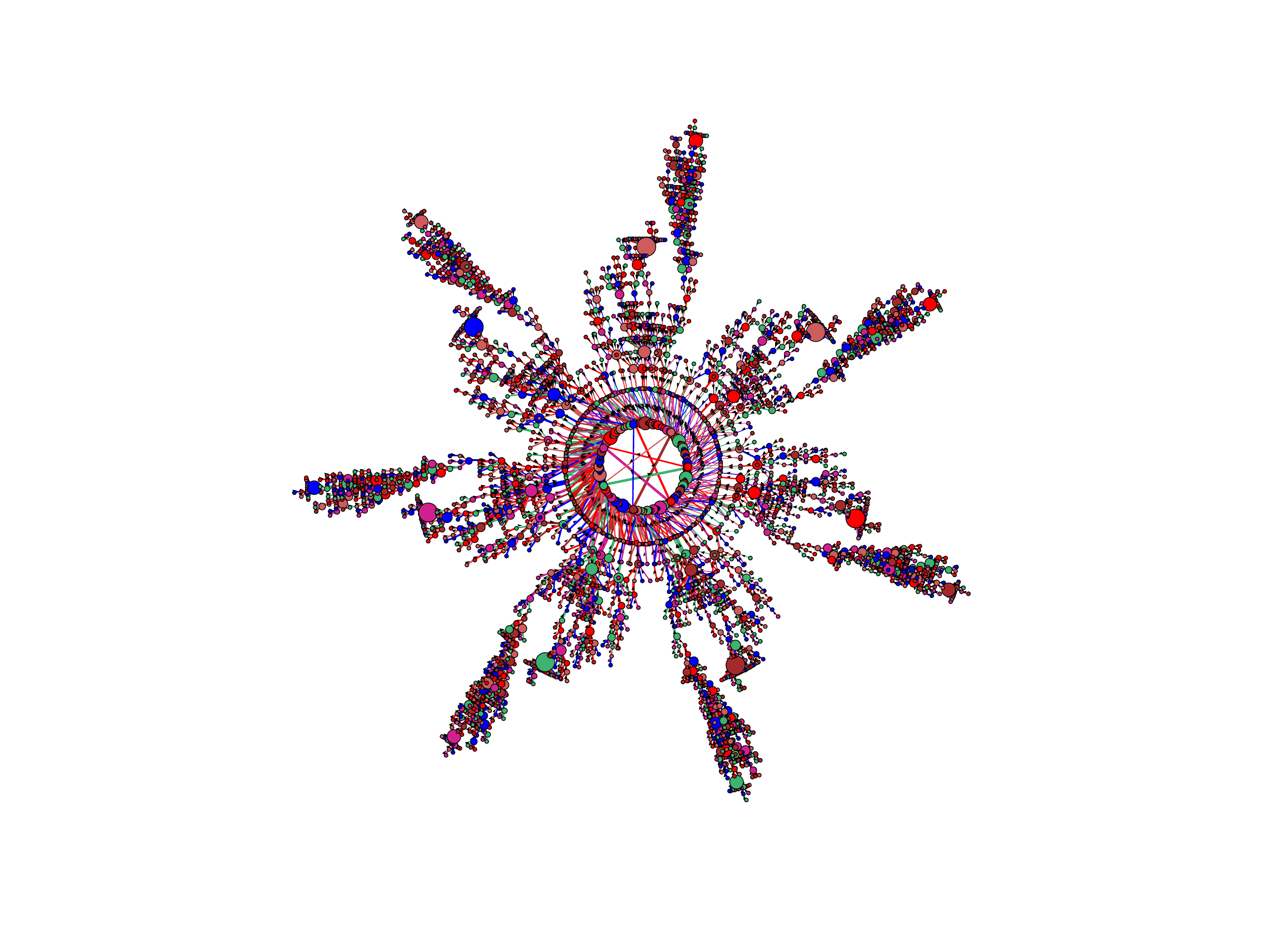}
       \end{minipage}\\[-4ex]
       \begin{center}           
       \textsf{\small (a) rcode(hex)76b5, seed(dec)=3187, attractor period period 140, states/nodes 4333. }
       \end{center}       
%nextrow---------------------------------------------
       \begin{minipage}[b]{1\linewidth}
         \includegraphics[width=.48\linewidth,viewport=263 112 911 662,clip]{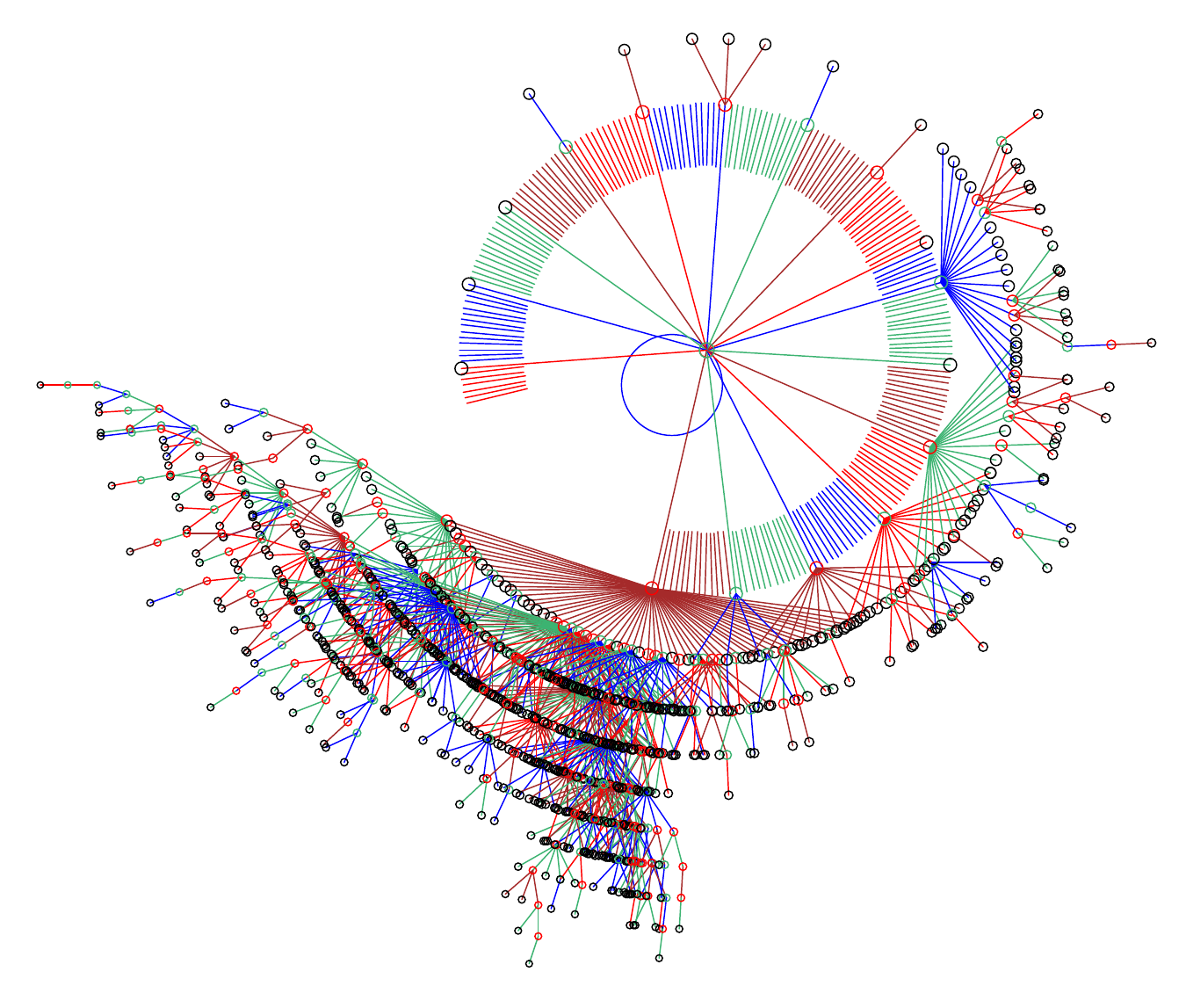}
     \hfill
       \includegraphics[width=.48\linewidth,viewport=263 112 911 662,clip]{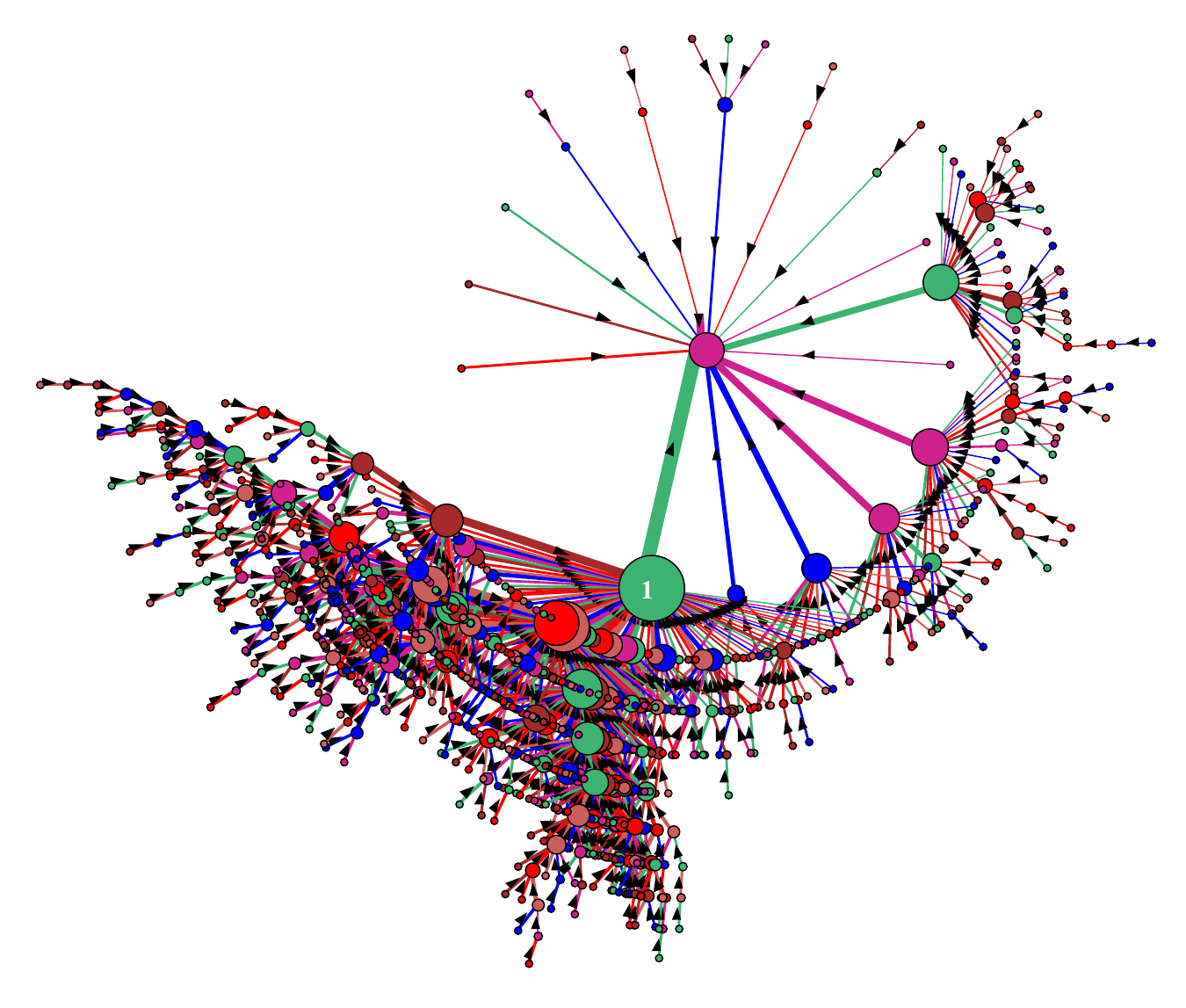}
       \end{minipage}\\[-5ex]
        \begin{center}   
          \textsf{\small (b) rcode (hex)ac88, point attractor and seed all0s,\\
          shown with equivalent subtrees suppressed, so 1132 states/nodes from 15541 in the basin}
        \end{center}
        \end{minipage}
        \end{center}   
       \vspace{-2ex}
       \caption[Examples of single basin, and its istr-graph]
               {\textsf{Examples of compressed single basins of attraction, 1d CA $v2k4$, $n$=14.
                   \underline{\it Left side}: the classic-graph,
                   \underline{\it Right side}: the default istr-graph with scaled nodes/links.
         \label{fig:quick_single_basin.ps}}}
         
\end{figure}

\begin{figure}[htb]
             \begin{center}
              \textsf{\small
             \begin{minipage}[t]{.38\linewidth}
               \includegraphics[width=1\linewidth]{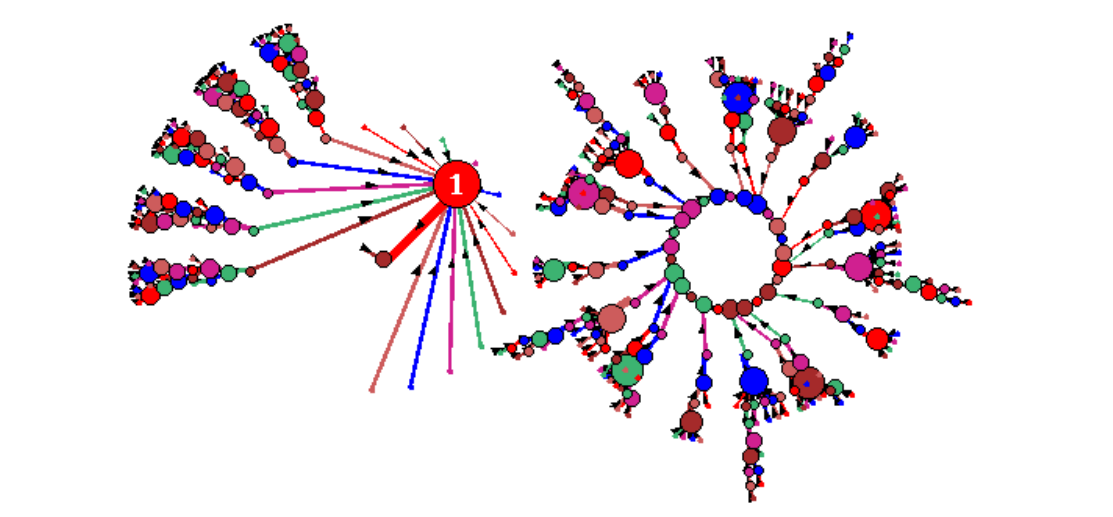}\\[-5ex]
               \begin{center}istr-graph with 2 basins after pause\end{center}
             \end{minipage}
             \hfill
             \begin{minipage}[b]{.58\linewidth}
               \vspace{-6ex}
               \includegraphics[width=1\linewidth]{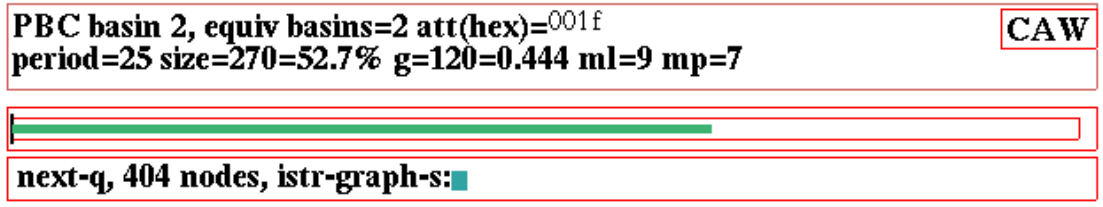}\\[-5ex]
              \begin{center}info, paused progress bar and prompt\\
                             ($v2k3$ $n$=10 rcode 110) \end{center}
             \end{minipage}             
           }
           \end{center}
           \vspace{-3ex}    
           \caption[istr-graph for a basin pause]
                   {\textsf{An example of the istr-graph of a 1d CA paused  basin of attraction field.
                       \underline{\it Right}: prompt to create the istr-graph, showing number of nodes.
                       \underline{\it Left}: the istr-graph.
                       \label{basin-pause-istr.ps}}}
           \vspace{10ex}        
\end{figure}

\begin{figure}[htb]
             \begin{center}
              \textsf{\small
             \begin{minipage}[t]{.38\linewidth}
               \includegraphics[width=1\linewidth,viewport=15 548 723 878, clip]{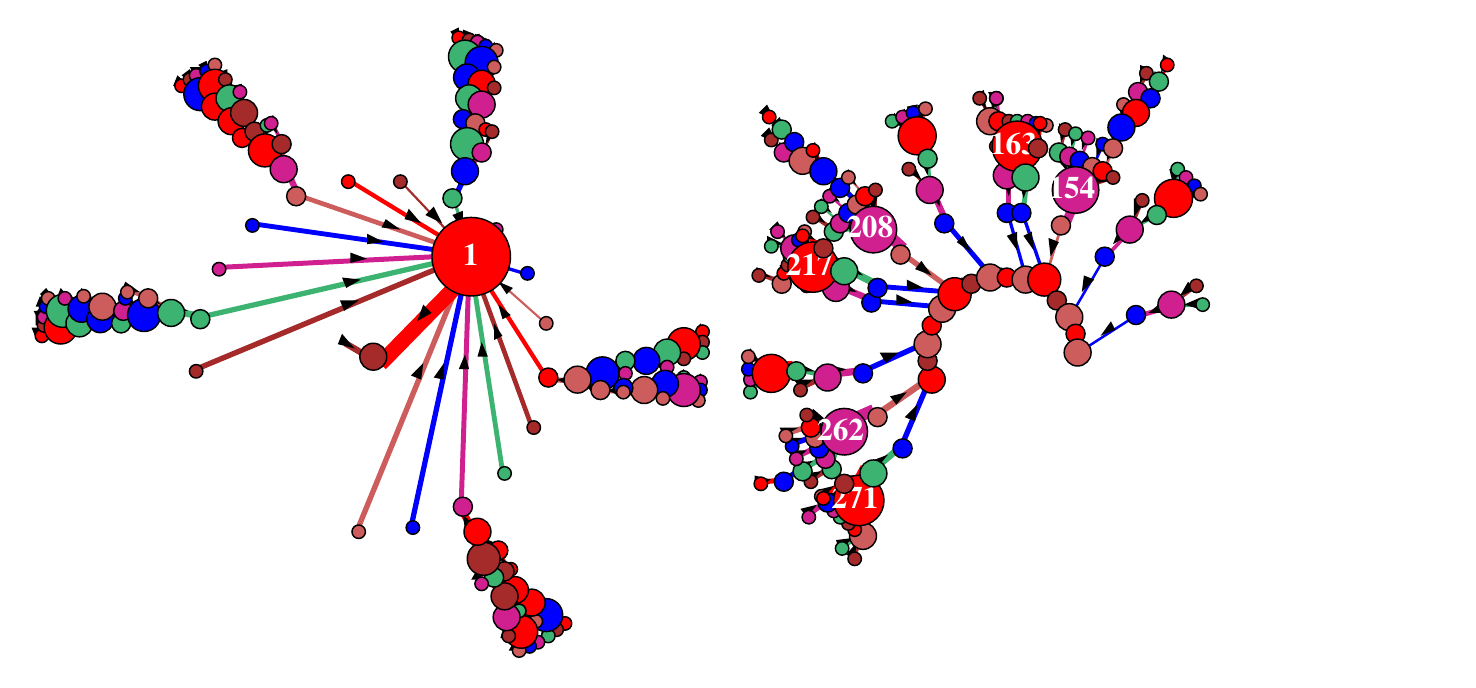}\\[-5ex]
               \begin{center}(a) istr-graph with incomplete field\end{center}
             \end{minipage}
             \hfill
             \begin{minipage}[b]{.58\linewidth}
               \vspace{-6ex}
                \includegraphics[width=1\linewidth]{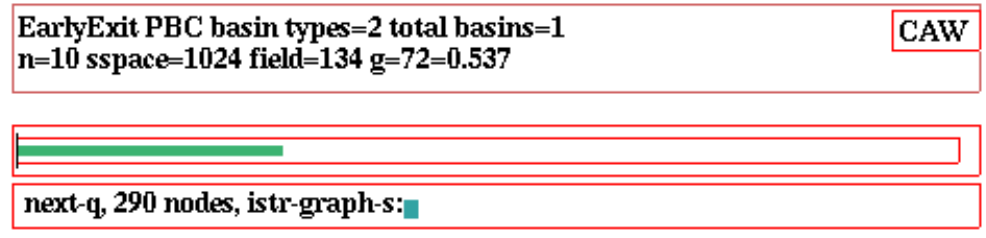}\\[-5ex]
              \begin{center}(a1) info, interrupted progress bar and prompt\\
                               ($v2k3$ $n$=10 rcode 110) \end{center}
             \end{minipage}
             %----------------------------------------
             \begin{minipage}[t]{.38\linewidth}
               \includegraphics[width=1\linewidth,viewport=60 222 1000 850, clip]{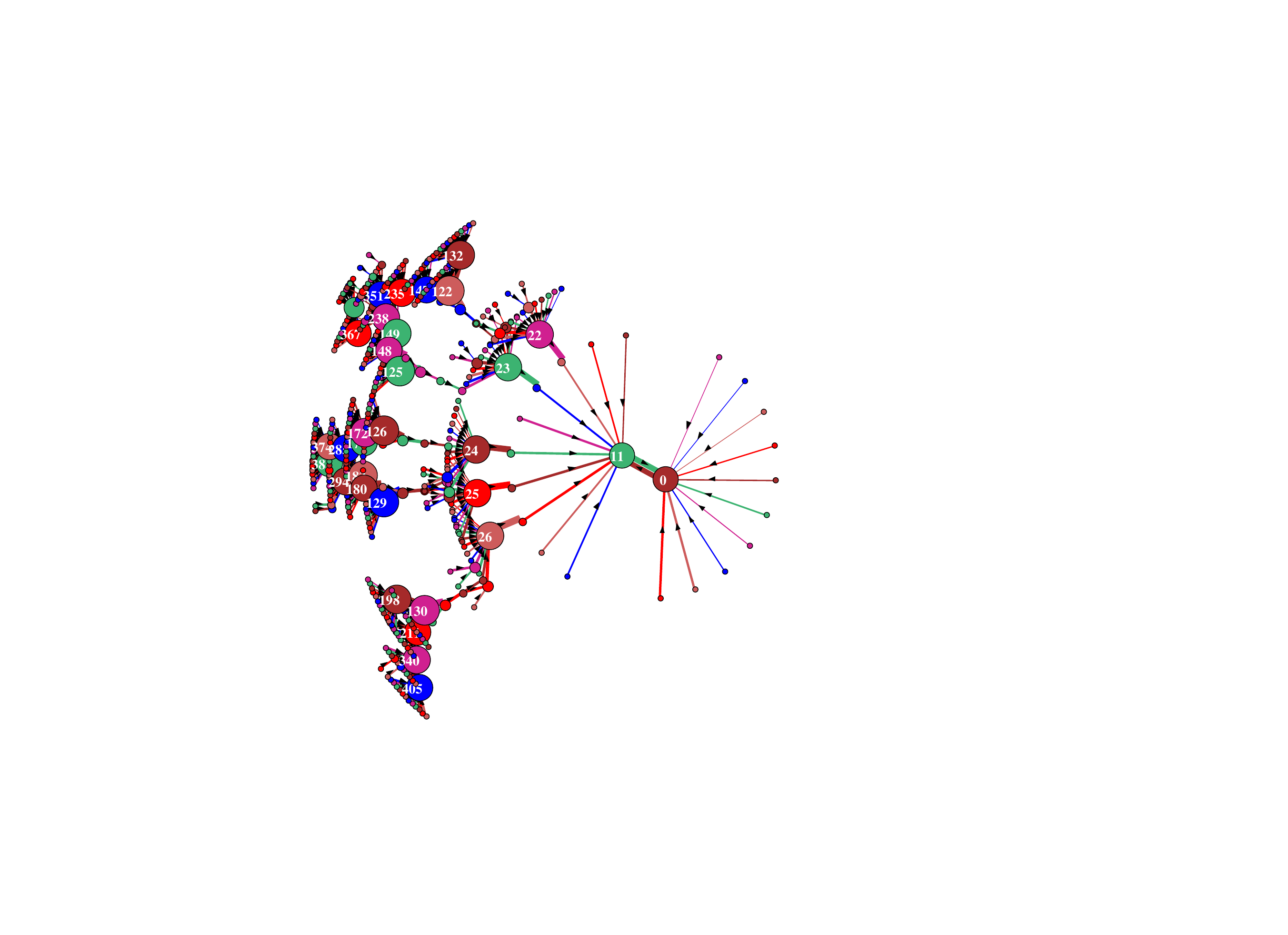}\\[-5ex]
               \begin{center}(b) istr-graph showing incomplete basin\end{center}
             \end{minipage}
             \hfill
             \begin{minipage}[b]{.58\linewidth}
               \vspace{-6ex}
               \includegraphics[width=1\linewidth]{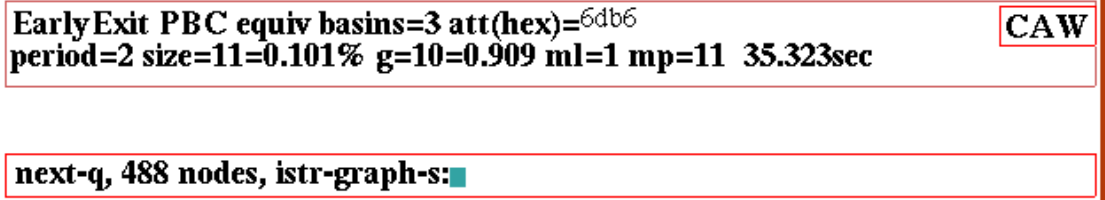}\\[-5ex]
              \begin{center}(b1) info, and prompt\\
                                 ($v2k5$ $n$=15 seed(hex)=6db6 tcode=53) \end{center}
             \end{minipage}\\[5ex]
             %----------------------------------------
             \begin{minipage}[t]{.38\linewidth}
               \includegraphics[width=1\linewidth,viewport=50 152 1200 826, clip]{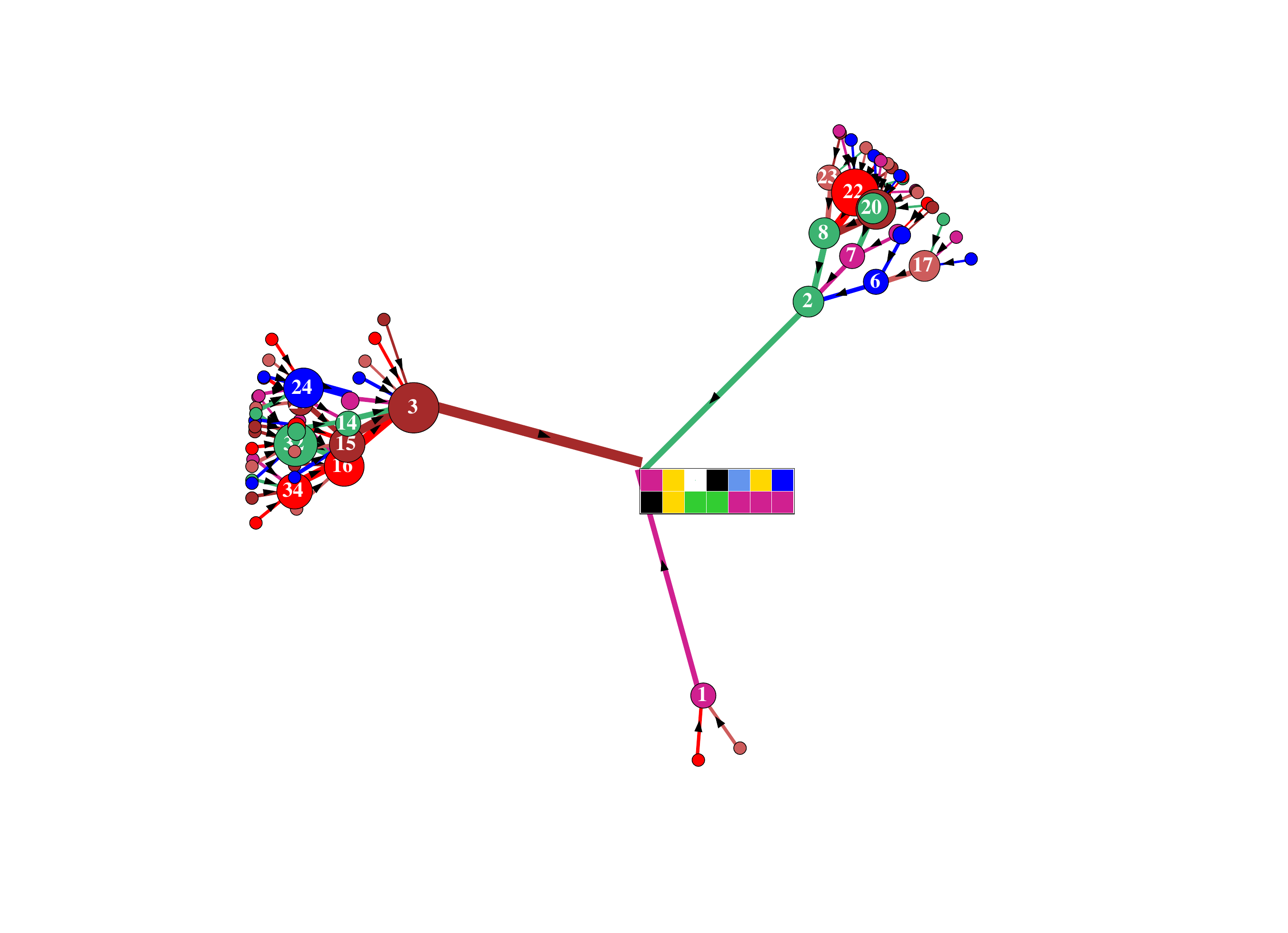}\\[-5ex]
               \begin{center}(c) istr-graph showing incomplete subtree, root shown as 7$\times$2 pattern\end{center}
             \end{minipage}
             \hfill
             \begin{minipage}[b]{.58\linewidth}
               \vspace{-6ex}
               \includegraphics[width=1\linewidth]{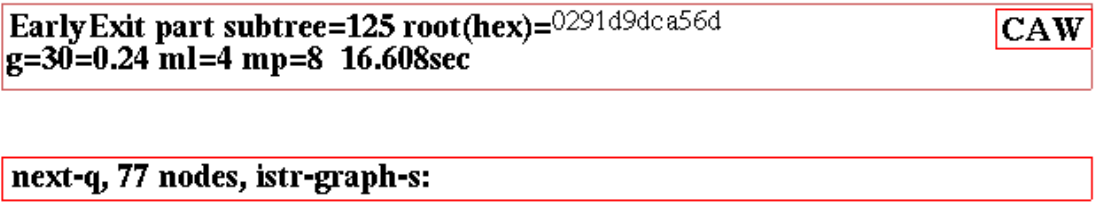}\\[-5ex]
              \begin{center}(c1) info, and prompt\\
                value range 8, random rcode and initial state\\
                forward 4 steps before backward\\
                ($v8k3$ $n$=14 :root(hex)= shown in info) \end{center}
             \end{minipage}
           }
           \end{center}
           \vspace{-2ex}    
           \caption[istr-graph for EarlyExit of classic-graph]
                   {\textsf{Examples of partly drawn istr-graphs for 1d CA
                        from interrupted (uncompressed) classic-graphs:
                       (a) basin of attraction field, (b) single basin, (c) subtree.
                       \underline{\it Right}: prompt to create the istr-graph, showing number of nodes.
                       \underline{\it Left}: the istr-graph.
                       Note that to reproduce these results,
                       drawing the classic-basins may need to be slowed
                       down\cite[\hspace{-1ex}\footnotesize{\#24.11}]{EDD}
                       to allow time to EarlyExit.
                       \label{EaryExit110-istr.ps}}}
           \end{figure}

\begin{figure}[htb]
\vspace*{-1ex}
\includegraphics[width=.85\linewidth,viewport=48 565 890 846,clip]{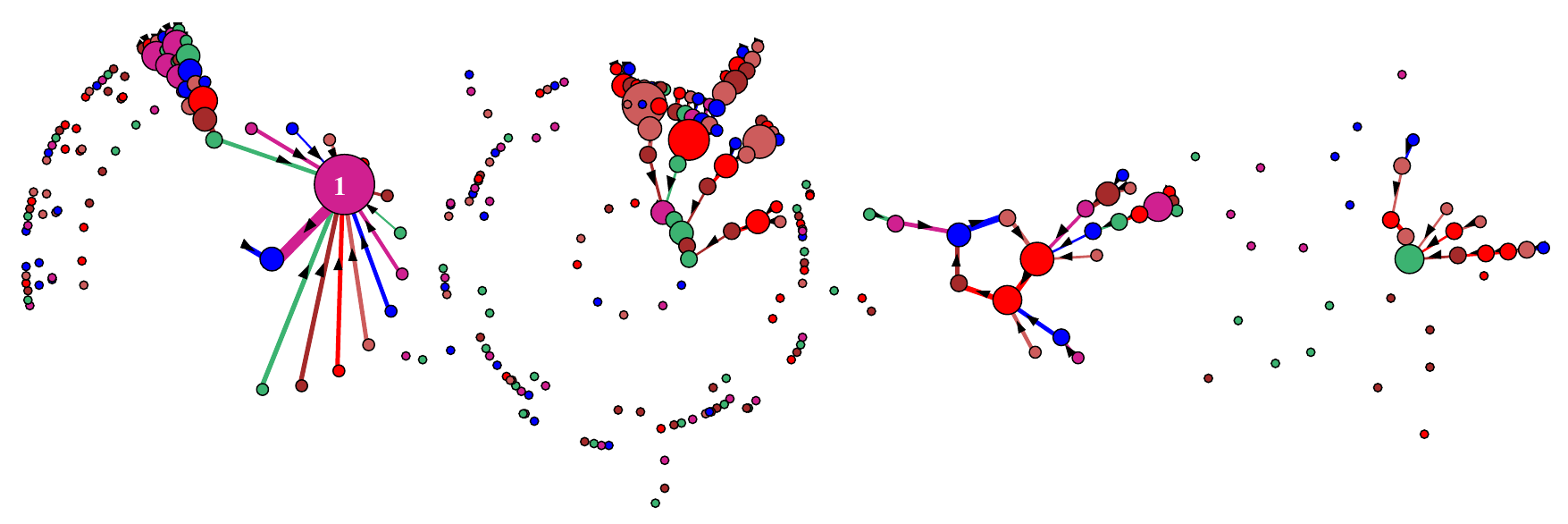}
\vspace*{-2ex}
\caption[Garden-of Eden states disconnected in istr-graph]
{\textsf{The istr-graph with default compression on, suppressed equivalent basins and trees,
but showing disconnected garden-of-Eden nodes in equivalent trees,
respecting options in \cite[\hspace{-1ex}\footnotesize{\#26.2.3}]{EDD}.
$v2k3$ rcode(dec)110, $n$=10.
}} \label{r110-GofE-istr.ps}
\end{figure}

\begin{figure}[htb]
\vspace*{-1ex}
%\fbox{
\begin{minipage}[b]{1\linewidth}
\includegraphics[width=1\linewidth,viewport=252 138 1310 780,clip]{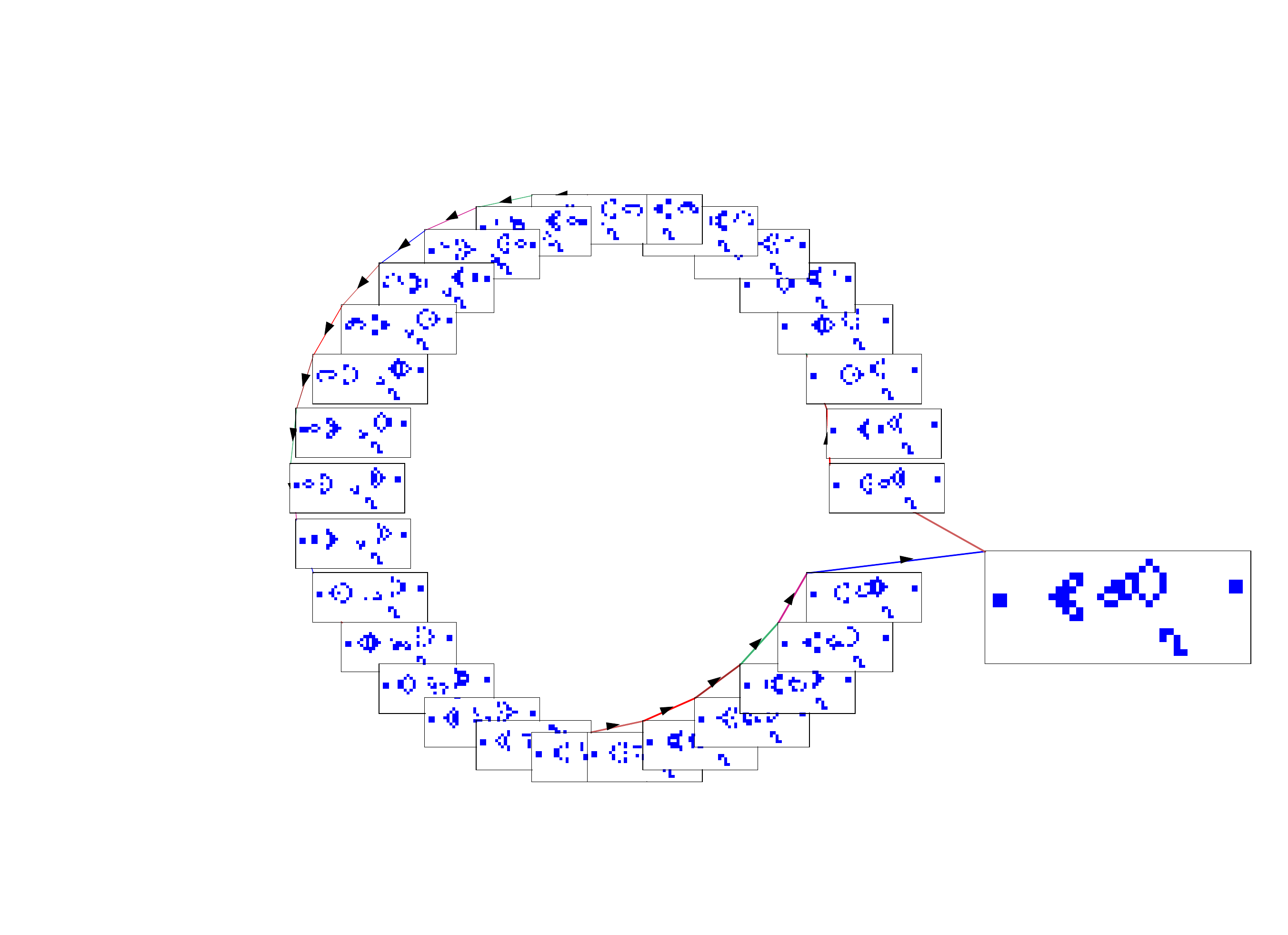}
\end{minipage}%}
\begin{flushright}
%\vspace*{-53.7ex}
\vspace*{-52.5ex}
\begin{minipage}[b]{.30\linewidth}
\caption[Game-of-Life glider-gun attractor]
{\textsf{
Gosper's glider-gun from the game-of-Life~\cite{conway}
with the glider stream absorbed by an eater,
shown as an istr-graph bare attractor with period 30,
based on the classic-graph with backward time-steps limited to zero.
Nodes shown in 2d (38$\times$16).
%as shown in figure~\mbox{\cite[\hspace{-1ex}\footnotesize{\#29.2}]{EDD}}.\phantom{xxxxx}
\phantom{xxxxxxxxxxxxxxxxxxxxxxxxxxxxxxxxxxxxxxxxxxxxxxxxxxxxxxxxxxx}
\phantom{xxxxxxxxxxxxxxxxxxxxxxxxxxxxxxxxxxxxxxxxxxxxxxxxxxxxxxxxxxx}
\phantom{xxxxxxxxxxxxxxxxxxxxxxxxxxxxxxxxxxxxxxxxxxxxxxxxxxxxxxxxxxx}
\phantom{xxxxxxxxxxxxxxxxxxxxxxxxxxxxxxxxxxxxxxxxxxxxxxxxxxxxxxxxxxx}
\phantom{xxxxxxxxxxxxxxxxxxxxxxxxxxxxxxxxxxxxxxxxxxxxxxxxxxxxxxxxxxx}
%Nodes shown in 2d (38$\times$16).
The initial state (above)
dragged out and expanded.
\label{Gosper38x16-istr.ps}} }
\end{minipage}
\end{flushright}
\end{figure}

\subsection{istr-graph compression of 1d CA dynamics}
\label{istr-graph compression of 1d CA dynamics}

The istr-graph will comply with compression of 1d CA dynamics,
and presents links between states in a compressed attractor cycle more
accurately than the classic-graph.

The order of states on the cycle of the uncompressed and compressed graphs
may be different, for example
figures~\ref{uncompressed and compressed istr-graph} (a)-(d), (b)-(e) and (c)-(f).
For the istr-graph this will be evident by internal cross links,
figure~\ref{uncompressed and compressed istr-graph} (e) and (f),
but not in the classic-graph figure~\ref{uncompressed and compressed istr-graph} (d),
In the classic compressed graph, in DDLab and examples
in \cite{Wuensche92}, these internal cross-links are absent. Although the order
of nodes and their subtrees is correct, the attractor cycle itself is simplified as
an external polygon.

\subsection{Selecting and activating the istr/ibaf graph}
\label{Selecting and activating the istr/ibaf graph}

\noindent The ibaf-graph applies only in FIELD-mode for an uncompressed basin of attraction field,
and its setup includes extra steps (section~\ref{extra steps for the ibaf-graph})
for the exhaustive reverse algorithm.
The istr-graph is more general as it applies to a single basin or
subtree in SEED-mode as well as to a basin of attraction field in
FIELD-mode.

To select the istr/ibaf graph\footnote{See \cite[\hspace{-1ex}\footnotesize{\#4.3}]{EDD} for
a quick-start example to select an istr/ibaf graph starting from the first DDLab prompt.},
at the first
{\color{BrickRed}{\bf \setlength{\fboxsep}{.3ex} \fbox{basin parameters}}}
prompt\cite[\hspace{-1ex}\footnotesize{\#24.1}]{EDD}
enter ``{\it graph-}{\bf g}'' to go directly to the prompt for interactive graphs,
or arrive there by viewing the output parameters in sequence, or backtrack
from any of these later prompts (having made selections) back to
the first prompt to enter ``{\it graph-}{\bf g}''.

The following top-right prompt is presented,

\begin{quote}
     \underline{\it in SEED-mode, only the istr-graph is available}\\
     {\bf {\color{BrickRed}interactive graphs:}} {\bf state transition graph, istr-s:}
\end{quote}

\begin{quote}
     \underline{\it in FIELD-mode, all graphs based on the field are available}\\
     {\bf {\color{BrickRed}interactive graphs:}} {\bf istr-s ibaf-i, jump-j(no-edges+L):}
\end{quote}

Enter ``{\it istr-}{\bf s}'' for the istr-graph in either SEED-mode  or FIELD-mode,
were compression is active by default for 1d CA, but can be
deactivated\cite[\hspace{-1ex}\footnotesize{\#26.2.1}]{EDD}, or in FIELD-mode only
enter ``{\it ibaf-}{\bf i}'' (or ``{\it jump-}{\bf j}'')\cite[\hspace{-1ex}\footnotesize{\#20.7}]{EDD}).

%\enlargethispage{3ex}
There will be further prompts to set
{\color{BrickRed}{\bf mutation}}\cite[\hspace{-1ex}\footnotesize{\#28}]{EDD}
%(chaper~\ref{Mutation of attractor basins})
--- ``{\it rcode-}{\bf r}'' is a good choice here for CA,
or ``{\bf return}'' for the default.\\

\noindent
\setlength{\fboxsep}{1ex}
\fbox{%this is a subsection in a frame
 \begin{minipage}{0.96\linewidth}%
 
\subsubsection{extra steps for the ibaf-graph}
\label{extra steps for the ibaf-graph}

If ``{\it ibaf-}{\bf i}'' was selected, ``compression'' usually on by default for 1d CA,
will be turned off automatically with the added  message {\bf $\dots$-compression OFF:}.
Proceed with these extra steps for the ibaf-graph,

\begin{s_itemize}
\setlength\itemsep{1ex}

\item Enter {\bf return} until the final ``{\bf basin field}'' prompt containing ...\hfill\break
``\mbox{{\bf \color{BlueViolet}{ibaf: exh-e rmap-r seq-s}}}''\cite[\hspace{-1ex}\footnotesize{\#29.4}]{EDD}.
The options that are valid for the ibaf-graph appear in blue (and underlined just here below for emphasis),

\begin{quote}
   {\bf basin field (local) n=10, nonlocal-x \underline{\color{BlueViolet}{ibaf: exh-e rmap-r seq-s}}
          \color{black}{nto-o}}\\
   {\bf view ppstack-1:} 
\end{quote}

Enter ``{\it exh-}{\bf e}'' for ``exhaustive pairs''\cite[\hspace{-1ex}\footnotesize{\#29.7.1}]{EDD}
--- ``{\it rmap-}{\bf r}'' for a random map or ``{\it seq-}{\bf s}'' for sequential updating are also
valid for their respective ibaf-graphs.

\item Then the following prompt is presented,

\begin{quote}
   {\bf exhaustive pairs: compute-ret, and save-s, and prt-p/+p:}
\end{quote}

Enter ``{\bf return}'' to compute the exhaustive list. See \cite[\hspace{-1ex}\footnotesize{\#29.7}]{EDD}
for a description of this and the other options).

\item A red progress bar will appear\cite[\hspace{-1ex}\footnotesize{\#29.7.1}]{EDD},
lasting a brief moment for a small state-space, but taking an  exponentially longer time for larger state-spaces,
thus the option \mbox{``{\it interrupt-}{\bf q}''},

\begin{quote}
  \hspace*{-1ex}\includegraphics[width=.8\linewidth]{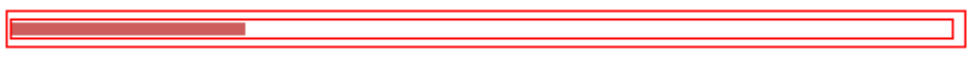}\\[-1ex]
   {\bf setting up exhaustive pairs, interrupt-q:}
\end{quote}
\end{s_itemize}

\end{minipage}
}

\subsubsection{istr/ibaf graph final prompt}
\label{istr/ibaf graph final prompt}

The classic-graph is drawn as usual in the main window ---
a green bar monitors progress\cite[\hspace{-1ex}\footnotesize{\#30.1}]{EDD}.
Once complete (or paused/interrupted, section~\ref{istr-graph on interrupt or pause})
a top-right data window appears above the bar, with the final prompt below the bar
to select the istr/ibaf graph (figure~\ref{istr/ibaf final prompt fig}) below.\\

\begin{figure}[H]
\vspace*{-2ex}
\begin{center}
\small
\raisebox{2ex}{(a)} \includegraphics[width=.6\linewidth]{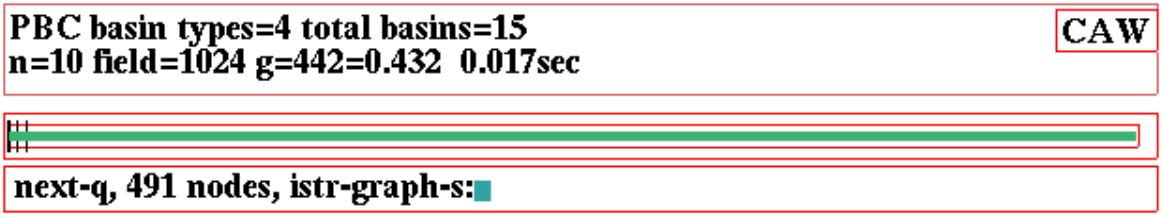}\\[3ex]
\raisebox{2ex}{(b)} \includegraphics[width=.6\linewidth]{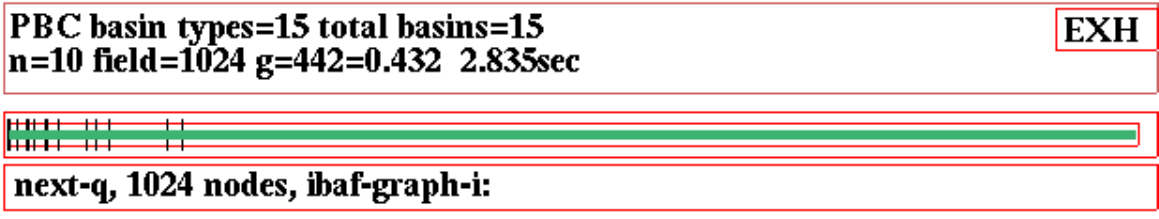}
\end{center}
\vspace{-4ex}
\normalsize
\caption[istr/ibaf activate]
{\textsf{The final prompt to activate: (a) the compressed istr-graph, and (b) the
uncompressed \mbox{ibaf-graph}, or continue to next mutant ---
for the same system and noting the number of nodes ($v2k3$, $n$=10 rcode 110).
}}
\label{istr/ibaf final prompt fig}
\end{figure}

\noindent
\setlength{\fboxsep}{1ex}
\fbox{%this is a subsection in a frame
 \begin{minipage}{0.96\linewidth}%

\subsubsection{istr-graph on interrupt or pause}
\label{istr-graph on interrupt or pause}
 
Drawing a classic-graph (field, single basin or subtree) can be
interrupted with ``{\bf q}'' and continued with ``
{\bf return}''\cite[\hspace{-1ex}\footnotesize{\#30.2}]{EDD},
but if instead ``{\bf q}'' is entered a second time to abandon the
classic-graph\cite[\hspace{-1ex}\footnotesize{\#30.2.3}]{EDD}
the istr-graph (if active) can be drawn at this point based on the partly
drawn classic-graph\cite[\hspace{-1ex}\footnotesize{\#30.2.4}]{EDD},
but this is a one-time opportunity.

\hspace{3ex} In a similar way, the classic-graph can be set to pause
(and continue) at various stages of detail\cite[\hspace{-1ex}\footnotesize{\#27.1.2}]{EDD},
and an istr-graph (but not an ibaf-graph) can be drawn after each basin is
complete\cite[\hspace{-1ex}\footnotesize{\#27.1.4}]{EDD}.
\end{minipage}
}\\[2ex]

Entering ``{\bf return}'' or ``{\bf q}'' will bring up the top-left
``Attractor basin complete prompt'' with its many options\footnote{Including ``{\bf speed(slow)-V}''\cite[\hspace{-1ex}\footnotesize{\#24.11}]{EDD}
  to allow enough time to test interrupt/pause.}\cite[\hspace{-1ex}\footnotesize{\#30.4}]{EDD}
but where ``{\bf return}''
will draw the next classic-graph according to mutation settings\cite[\hspace{-1ex}\footnotesize{\#28}]{EDD},
then recycle to the istr/ibaf final prompt (figure~\ref{istr/ibaf final prompt fig}), which can be repeated
indefinitely.

Enter ``{\it istr-}{\bf s}'' or ``{\it ibaf-}{\bf i}''  at the final prompt
(section~\ref{istr/ibaf graph final prompt} to compute and draw the interactive graph.
The computation is tracked showing the percentage computed so far, for example, 

\begin{quote}
  \mbox{{\bf computing istr-graph, 10971 nodes, quit-q, or wait...
                    \setlength{\fboxsep}{.2ex} \fcolorbox{BrickRed}{white}{14\%}}
                               {\it \small 1d CA $v2k3$ $n$=14}}\\ 
  {\bf computing ibaf-graph, 16384 nodes, quit-q, or wait...
                     \setlength{\fboxsep}{.2ex} \fcolorbox{BrickRed}{white}{5\%}}
                                     {\it \small (rcode 45,  as above)}
\end{quote}

Enter ``{\bf q}'' to abandon if the wait is too long,
and ``{\bf return}'' for the next mutant, and the cycle will repeat.

The time to compute the istr/ibaf graph depends on the number of nodes
(figure~\ref{istr/ibaf final prompt fig}
independent of $k$ or the wiring/rule architecture. For the ibaf-graph the time is especially
sensitive to the size of state-space,
but for the istr-graph there may well be fewer nodes,
because of compression, single basins. subtrees, limiting the number of levels,
and the possibility to interrupt/pause for a partly completed istr-graph.

Once computed, the istr/ibaf graph is drawn as in
figure~\ref{r110ibaf.ps}(b,c) for example.
At the same time, the top-right initial reminder
(section~\ref{The istr/ibaf graph initial reminder}) is presented.

\section{Interactive graph reminders}
\label{Interactive graph reminders}

\noindent Options for the interactive graphs are
presented at two interchangeable stages with distinct top-right
reminders.
The initial-reminder (section~\ref{The istr/ibaf graph initial reminder})
applies to the whole graph --- all nodes simultaneously.
The drag-reminder (section~\ref{The istr/ibaf graph drag reminder})
can also apply to the whole,
but options generally apply to a single node, a geometric range of
nodes (blocks), or most significantly to the selected node and its
connected fragment. 
Section~\ref{Common themes between initial and drag graphs}
covers common themes;
sections~\ref{Initial-graph options} and\ref{Drag-graph options}
give lists of the initial and drag options. 
The reminders are very similar across all graph types,
but special cases are noted in these lists.

Enter ``{\bf q}'' to quit the initial graph/reminder and backtrack
to the top-left
``Attractor basin complete prompt''\cite[\hspace{-1ex}\footnotesize{\#30.4}]{EDD}.
Move from initial to drag reminder with a mouse click, and from drag back to the
initial reminder with ``{\bf q}''.

\subsection{The istr/ibaf graph initial reminder}
\label{The istr/ibaf graph initial reminder}

The initial reminder for the istr/ibaf graphs is shown below
 --- the network-graph and jump-graph reminders are similar.

 \begin{quote}
     \underline{\it initial-reminder applies to whole graph --- this example for the istr/ibaf graph}\\
     \mbox{{\bf {\color{BrickRed}ISTR-graph:}} {\bf drag-(def) PScript-P net=\# ant-a unscram-u win-w rank0-k}}\\
     {\bf settings-S rot-x/X flip-h/v nodes-(/)= links-\{/\} both-[/]}\\% as: exp/contr:nodes-e/c links-E/C both-B/b}
     {\bf Unreach-U matrix-t/T nodes-n/N links-l Labels-+ arrows-A/}{$\boldsymbol <$}{\bf /}{$\boldsymbol >$}\\
     \mbox{{\bf layout: file-f graph-g circle/spiral-o/O 1d/2d(tog)/3d-1/2/3 rnd-r/R quit-q:}}\\
or\\
     {\bf {\color{BrickRed}IBAF-graph:} $\dots$}
     {\it \small (the rest as above)}
\end{quote}

\iffalse\footnotetext{To be meaningful,
      the adjacency-matrix\cite[\hspace{-1ex}\footnotesize{\#20.19}]{EDD}% (section~\ref{The adjacency-matrix})
      requires nodes numbered by decimal state as in the ibaf, network, and jump graphs. Nodes in the istr-graph
      are numbered by the order of computation, so although the option is nevertheless applicable, it will yield
      trivial results.}
\fi

The initial istr-graph (and ibaf-graph) layout will follow that of the classic-graph, as a subtree, single basin, 
or a basin of attraction field, separated into independent components,
but the layout would be disrupted by the options:
  \begin{center}''{\bf unscram-u}'', ``{\bf rank0-k}'',
``{\bf circle/spiral-o/O}'', \mbox{``{\bf 1d/2d(tog)/3d-1/2/3}''},
  ``{\bf rnd-r/R}''\end{center}
However these options are of some interest for a field because they demonstrate how
components can be dragged out from overlapping layouts.
The initial layout can be restored at any time with ``{\bf graph-g}''.

See section~\ref{Initial-graph options} for the initial decode
list\mbox{\cite[\hspace{-1ex}\footnotesize{\#20.9}]{EDD}}.
Enter ``{\bf q}'' to quit the initial-graph (and ``{\bf return}'' for the next mutant),
or ``{\bf q}' again to backtrack.

Click the left or right mouse button with the pointer anywhere on the
screen (or press {\bf return}) to change to the drag-graph and its
drag-reminder that applies to a selected node (click to
activate). For the istr/ibaf graphs the initial setting is 
``{\color{BrickRed}{\bf either, step=nolimit:}}'' 
for the node's linked fragment.% (section~\ref{drag graph: linked fragment in/out/either}).
For the network or jump graphs the
initial setting is ``{\color{BrickRed}{\bf single:}}''
for single node status.%(section~\ref{drag graph: single node}).

\subsection{The istr/ibaf graph drag reminder}
\label{The istr/ibaf graph drag reminder}

\noindent Enter {\bf return} or a left-mouse click in the initial-graph
to switch to the drag graph and top-right drag reminder below.
Left/right click to activate a node, and drag with the left mouse button depressed.
The default is to drag a ``linked fragment''
indicated by ``{\bf \color{BrickRed}{\bf either, step=nolimit:}}'',
which means that dragging the current active node will also drag its linked component ---
the entire basin of attraction.
%The decode list is in section~\ref{Drag-graph options}.%examples for istr-graph v3k3 n=6

\begin{quote}
     \underline{\it drag-reminder applies to a node and its fragment, in this case the entire basin}\\
    \mbox{{\color{BrickRed}{\bf node 25, either, step=nolimit:}} {\bf leftb-drag  PScript-P elstc/snap-d gap-g}}\\
     \mbox{\bf inactive?-rightb first, rot-x/X flip-h/v nodes-(/)/=/E links-\{/\} both-[/] just-j/J}\\
     {\bf Lnk25:cut/restore-c/r Lnks25-28:cut/add/restore-C/A/R net-\#}\\
     {\bf step-(1-9) nolimit-0 single-s in/out/either-i/o/e all-a exit-q:}
\end{quote}

The type of linked fragment is changed with ``{\bf in/out/either-i/o/e}'', and the time-step distance
``{\bf step-(1-9)}'' or ``{\bf nolimit-0}''. Links can also be cut/added.
Enter ``{\it single-}{\bf s}'' to change to single node status,
which allows labels and blocks, %(sections~\ref{drag graph: block}, \ref{Defining a block})
though blocks are not so useful for the istr/ibaf graph.

\begin{quote}
      \underline{\it ``single'' istr-graph drag reminder, for example}\\
    \mbox{{\color{BrickRed}{\bf node 25, single:}} {\bf leftb-drag PScript-P elstc/snap-d gap-g}}\\
     \mbox{\bf inactive?-rightb first, nodes-(/)/= just-j/J}\\
     {\bf block-B Label-L/+ net-\#}\\
     {\bf step-(1-9) nolimit-0 single-s in/out/either-i/o/e all-a exit-q:}
\end{quote}

See section~\ref{Drag-graph options} for the drag decode
list\mbox{\cite[\hspace{-1ex}\footnotesize{\#20.10}]{EDD}}.
Enter ``{\bf q}'' to exit the drag-graph and switch back to the initial-graph.

%--------------------------------------------------------------------------
\section{Common themes between initial and drag graphs}
\label{Common themes between initial and drag graphs}

\noindent By default, nodes (vertices) are displayed as colored discs cycling
through four colors.
There are two designs of centrally placed disc numbers, or numbers can be omitted,
operating on a 3-way toggle~``{\bf nodes-N}'' at the initial options,
with the current design inherited by the drag-graph.
The default is a white number on the colored disc if the disc is big enough
\raisebox{-0.5ex}[0pt][0pt]{\includegraphics[height=.03\linewidth,viewport=734 501 816 543,clip]{r30n10s971-istr}}
(i.e.~figure~\ref{node-dis.ps}).
The alternative is a number on a white rectangle irrespective of disc size so always
visible, which may fit within its disc or cover it
\raisebox{-0.5ex}[0pt][0pt]{\includegraphics[height=.025\linewidth, viewport=325 460 406 485,clip]{r110istr-basin1-nframe}}
(i.e~figure~\ref{r110istr-basin1-nframe.ps}) ---the color of the number and its frame follows the disc color.
Links (edges) are colored according to the parent node and aligned
asymmetrically anti-clockwise relative to the parent to separate mutual
link overlap by a central gap. Self-links are shown as short stubs
projecting from a node.  Link direction is indicated by centrally
placed arrows, size adjustable, and colored black unless links are
thin lines; in this case the arrow color follows line color.

Disc numbers show the network
order, 0 to $n$-1 for the network-graph, 0 to the node total-1 for the
istr-graph following the sequence of node computation\footnote{The
istr-graph algorithm computes nodes in sequence which is reflected in
node/disc numbers (figures~\ref{node-dis.ps}, \ref{istr-graph single basin}),
whereas the ibaf-graph node/disc numbers follow the decimal
state.}, 0 to $v^k$-1 (state-space) for the ibaf-graph, and 1 to the
number of basins for the jump-graph.

Both the initial and drag graphs are displayed in a large central window,
and for both ``\mbox{\bf  PScript-P}'' will save the current image as a
vector PostScript file\mbox{\cite[\hspace{-1ex}\footnotesize{\#20.15}]{EDD}},
and ``{\bf net-\#}'' will
redrawn the graph, restoring any link cuts or additions
made in the drag-graph.\\

\subsection{Common presentation options}
\label{Common presentation options}

\noindent The following presentation options feature in both the initial-graph, and the drag-graph
if a fragment is active. 

\begin{center}
         {\bf rot-x/X flip-h/v nodes-(/)/= links-\{/\} both-[/]}\\
         {\it \small (only ``{\bf nodes-(/)/=}'' features for  single node status in the drag-graph)}\\
\end{center}

\begin{list}{$\Box$}{\parsep 0ex \itemsep .6ex  
 \leftmargin 20ex  \rightmargin 4ex \labelwidth 25ex \labelsep 1ex}
 \item[\underline{\it prompts} $\dots$]   \underline{\it what they do}
 \item[{\bf rot-x/X} $\dots$] lower-case ``{\bf x}'' rotates clockwise,
   upper-case ``{\bf X}'' rotates anti-clockwise.
\item[{\bf flip-h/v} $\dots$] ``{\bf h}'' flips horizontally, ``{\bf v}'' flips vertically, 
\item[{\bf nodes-(/)} $\dots$] simple brackets: ``{\bf (}'' to contract, ``{\bf )}''
  to expand the node display as discs, numbers, or patterns.  The font
  size of numbers in the initial-graph will also set the size font of
  disc numbers, which is inherited by the drag graphs.

\item[{\bf nodes-=} $\dots$] enter ``{\bf =}'' (equals sign) to cycle through successive
                             node displays (figure~\ref{node-dis.ps}) depending
                             on the graph type as follows,
                             \begin{list}{$\Box$}{\parsep 0ex \itemsep 0ex  
                              \leftmargin 25ex  \rightmargin 4ex \labelwidth 45ex \labelsep 1ex}
                 \item[istr/ibaf graph: $\dots$] 6-way, discs/dec/hex/1d/2d/empty.
                 \item[linked network/jump graph: $\dots$] 3-way discs/decimal/empty
                 \item[unlinked network/jump graph: $\dots$] 2-way discs/decimal.
                              \end{list}
\item[{\bf links-\{/\}} $\dots$] curly brackets: ``{\bf \{}'' to contract,
           ``{\bf \}}'' to expand the length of links, or the distance between nodes.
                 Note that scaled link thickness follows disc size.
                 
\item[{\bf both-[/]} $\dots$] square brackets:  ``{\bf [}'' to contract,
              ``{\bf ]}'' to expand, both nodes and links together.

\end{list}

The  operations for  rotations, flips, and contract/expand link length, are made with reference to
the window center for the initial-graph, and the pointer position for the drag-graph.\\

%-------------------------------------------------------------------
\section{Initial-graph options}
\label{Initial-graph options}

\noindent A list of the options in the network/istr/ibaf/jump-graph initial reminder 
(section~\ref{The istr/ibaf graph initial reminder})
is set out and decoded in section~\ref{initial-options decode list}
in roughly the reminder order.
They apply to all nodes simultaneously,
though this is also true for the drag-graph with its option ``{\bf all-a}''.
Scaled discs/links is the initial default.
Titles in red give the graph type,

\begin{quote}
{\color{BrickRed}{\bf NET-graph:}}, {\color{BrickRed}{\bf NET-graph(nolinks):}}\\
{\color{BrickRed}{\bf ISTR-graph:}},  {\color{BrickRed}{\bf IBAF-graph:}}\\
{\color{BrickRed}{\bf JUMP-graph:}}, {\color{BrickRed}{\bf JUMP-graph(noilnks):}}, {\color{BrickRed}{\bf JUMP-graph(hist):}}
\end{quote}

Key-hits usually take effect without entering {\bf return}, but in some cases there
are suboptions as noted. Options that require a linked graph are indicated by
``\mbox{({\it linked only})}''.\\

\subsection{initial-options decode list}
\label{initial-options decode list}

\begin{list}{$\Box$}{\parsep 0ex \itemsep .6ex
    \leftmargin 15ex \rightmargin 0ex \labelwidth 30ex \labelsep 1ex}
\item[\underline{\it options} $\dots$]   \underline{\it what they mean} 
\item[{\bf drag-(def)} $\dots$] enter {\bf return}, or click the left or right
                             mouse button, for the ``drag''
                             options (section~\ref{Drag-graph options}). 
                             The drag-reminder will replace the initial-reminder
                             --- enter ``{\bf q}'' to revert.
                             Click on a node to activate it in the drag-graph.
\item[{\bf PScript-P} $\dots$] to save the current graph image as it appears --- a vector PostScript file
                       --- suboptions appear\cite[\hspace{-1ex}\footnotesize{\#20.15}]{EDD}.
\item[{\bf net-\#} $\dots$]  ({\it linked only}) to redraw the graph, restoring any link cuts or additions
                             made in the drag-graph.                              
\item[{\bf ant-a} $\dots$]   ({\it linked only}) to launch a projectile --- a probabilistic 
                             ``ant'' --- in the graph to trace a (Markov chain) path 
                             according to link probabilities, keeping track 
                             of the frequency of visiting vertices.
                             Top-right suboptions appear\cite[\hspace{-1ex}\footnotesize{\#20.13}]{EDD}.
                             %described in section~\ref{Probabilistic ``ant''}.
\item[{\bf unscram-u} $\dots$]  ({\it linked only}) to automatically unscramble the graph,
                             placing weakly connected nodes on the outer edges,
                             nearby their connections.
                             This works best in the circle or spiral layout.                
\item[{\bf win-w} $\dots$]  to show what lies ``below'' the graph window, which will include a top-right inset
                            \setlength{\fboxsep}{.2ex} \fcolorbox{BrickRed}{white}{\bf restore graph -any key:},
                            --- so toggling between.
                            For the istr/ibaf graphs and the {\it f}-jump graphs, to see the
                            precursor classic-graph
                            and its top-right data \mbox{window\cite[\hspace{-1ex}\footnotesize{\#27.2.1}]{EDD}}.
                            For the {\it h}-jump-graph to see the generating
                            \mbox{histogram\cite[\hspace{-1ex}\footnotesize{\#31.7.8}]{EDD}}.
                            For the network-graph initiated from the
                            ``wiring -graphic''\cite[\hspace{-1ex}\footnotesize{\#17.3}]{EDD} to compare
                            the two presentations\cite[\hspace{-1ex}\footnotesize{\#20.5.3}]{EDD}.
\enlargethispage{2ex}
\item[{\bf rank0-k $\dots$}] to toggle between default and ranked node positions.
                             {\bf rank0}/{\bf rank1} shows which is current.
                             The nodes are ranked (reordered) according to scaled node size, which
                             differs for graph type as follows,
                             
                             \begin{list}{$\Box$}{\parsep 0ex \itemsep 0ex  
                              \leftmargin 5ex  \rightmargin 0ex \labelwidth 15ex \labelsep 0ex}
                             \item[network-graph: $\dots$] ({\it linked only}) node size based on
                                inputs or outputs, whichever is active.                               
                              \item[istr/ibaf graph: $\dots$]
                                   node size based on in-degree (number of pre-images) so inputs.
                                A linear layout (i.e. circle or spiral)
                                disrupts the default basin layout, but becomes 
                                interesting if ranked because it reflects the
                                ``in-degree frequency histogram''\mbox{\cite[\hspace{-1ex}\footnotesize{\#24.6}]{EDD}}.
                              \item[jump-graph:$\dots$] node size based on basin volume.                      
                            \end{list}

\item[{\bf settings-S $\dots$}] to revise the default settings for the 
                             rotation angle, and the scaling factors for
                             nodes, links and arrows --- suboptions in
                             \cite[\hspace{-1ex}\footnotesize{\#20.17}]{EDD}.
\item[{\bf rot-x/X $\dots$}] to rotate the graph about the window center
                              by the default angle of 15 degrees, or a revised angle in ``{\it settings-}{\bf S}''
                              above.
                              Enter ``{\bf x}'' to rotate clockwise, or ``{\bf X}'' anti-clockwise. 
\item[{\bf flip-h/v $\dots$}] to flip the graph about the window center, enter ``{\bf h}'' to flip horizontally,
                              or ``{\bf v}'' to flip vertically .
\item[{\bf nodes-= $\dots$}] (equals sign) to cycle (toggle) though successive
                             node displays (figure~\ref{node-dis.ps}). The toggle sequence depends
                             on the graph type as in section~\ref{Common presentation options}.
                             
\item[{\bf nodes-(/) $\dots$}]  (simple brackets) enter 
                              ''{\bf (}'' to contract, or ``{\bf )}'' to expand,
                              the current node display\footnote{For discs (and 1d/2d patterns in
                              the istr/ibaf graph)
                              by the default factor (1.2), or a revised factor in ``{\it settings-}{\bf S}''.
                             For decimal/hexadecimal contract/expand by font pixel size between 6px and 150px,} 
                              and also equalise its size across the whole graph\footnote{Similar
                              to {\bf nodes-E} in the drag reminder to equalise node size in the fragment.}.  
                                                         
\item[{\bf links-\{/\}} $\dots$] (curly brackets) enter 
                     ``{\bf \{}'' to contract, or ``{\bf \}}'' to
                             expand the length of links (or distance between nodes for unlinked graphs)
                             by a default factor (1.1), or a revised factor in ``{\it settings-}{\bf S}''.
                
\item[{\bf both-[/]} $\dots$] (square brackets)
                             enter ``{\bf [}''to contract, or ``{\bf ]}''  to expand
                             both nodes and links/distances at the same time by the  
                             default factors or revised factors above.

\item[{\bf basins-i/I/s $\dots$}] ({\it f-jump-graph only}) enter ``{\it basins-}{\bf i}''
                             to redraw basins at jump-graph nodes but without the graph itself,
                             enter \mbox{``{\it basins-}{\bf I}''}  to keep the graph and draw
                             basins inside (or on top of) nodes\cite[\hspace{-1ex}\footnotesize{\#20.14}]{EDD}.
                             Enter ``{\it basins-}{\bf s}'' to change the basin scale ---
                             with suboptions\cite[\hspace{-1ex}\footnotesize{\#20.14}]{EDD}.
                             This is an alternative method for a flexible basin layout
                             with geometric possibilities for the ibaf-graph (but not for the istr-graph).
                             Once drawn, basin state-space coordinates can be saved, to reload
                             into the ibaf-graph\cite[\hspace{-1ex}\footnotesize{\#20.14.1}]{EDD}.

\item[{\bf Unreach-U $\dots$}] ({\it linked only})  to identify, disconnect and/or separate unreachable or 
                             hard to reach nodes in the graph. Suboptions appear\cite[\hspace{-1ex}\footnotesize{\#20.18}]{EDD}. 
\item[{\bf matrix-t/T $\dots$}] ({\it linked only}) to toggle between the graph
                            and its corresponding 2d ``jump-table'' or ``adjacency-matrix'' ---
                            top-right suboptions appear\cite[\hspace{-1ex}\footnotesize{\#20.19}]{EDD}.
                            For the ibaf-graph,  ``{\it matrix-}{\bf t}'' shows the matrix as an
                            unnumbered \mbox{pattern\cite[\hspace{-1ex}\footnotesize{\#20.19.3}]{EDD}} and
                            ``{\it matrix-}{\bf T}'' shows the list of exhaustive pairs in the
                            terminal\cite[\hspace{-1ex}\footnotesize{\#29.7.4}]{EDD}

                             \hspace{3ex} For the istr-graph, ``{\it matrix-}{\bf t}'' works as above, but as disc
                             numbers follow the order of computation
                             (not the decimal state) the result will be trivial. However, ``{\it matrix-}{\bf T}''
                             is useful as it shows a list: x to successor-of-x,  
                             for both disc numbers (in brackets)  and decimal states,
                             for example \texttt{(24>21) 40>57}.

\item[{\bf nodes-n $\dots$}] ({\it any graph type, but network-graph linked only})
                           to toggle between scaled and unscaled discs (and links).

\item[{\bf nodes-N $\dots$}] a 3-way toggle of the disc number presentation, which is inherited by the drag-graph.

        \begin{s_itemize}
        \setlength\itemsep{0ex}
      \item a white central number --- only appears if the disc is big enough, with a
                   diameter $\geq$ $1.4$ of the current text height (figure~\ref{fig:rbn_P}).   
            \item a central disc-colored number+frame with a white background,
                              always appears (figure~\ref{r110istr-basin1-nframe.ps}).
            \item a disc without a number.
        \end{s_itemize}
         The disc number size, which corresponds to the font
                            size set with ``{\it nodes-}{\bf (/)}'' above, is not affected.

\item[{\bf links-l $\dots$}] ({\it linked only}) to toggle between scaled links/edges and 
                          thin lines.
\item[{\bf Labels-+ $\dots$}] to toggle between  existing ``Labels'' and the current node display.
                            Labels\cite[\hspace{-1ex}\footnotesize{\#20.12}]{EDD}
                            can be created/deleted in the drag-graph.
\item[{\bf arrows-A/}{$\boldsymbol <$}{\bf /}{$\boldsymbol >$} $\dots$] 
                            ({\it linked only}) enter ``{\it arrows-}{\bf A}'' to toggle showing arrows.
                            Enter ``{\it arrows-}{$\boldsymbol <$}'' to decrease, or  ``{\it arrows-}{$\boldsymbol >$}''
                            to increase, the arrow size by 10\%.
                            The current arrow start size can be changed in ``{\it settings-}{\bf S}''.
\item[{\bf in/out-z $\dots$}]  ({\it linked network-graph only})
                            to toggle scaling based on inputs or outputs.
                            ``{\bf in/out}'' or ``{\bf out/in}'' shows which is current.
                            Nodes/links will be scaled if currently unscaled.
                            
\item[\underline{\it {\bf layout:} options} $\dots$]
\item[{\bf file-f $\dots$}] to save node coordinates of the current graph layout,
                             or load an existing file of the same system
                             --- a \texttt{.grh} file --- suboptions apply\cite[\hspace{-1ex}\footnotesize{\#20.16}]{EDD}.

\item[{\bf graph-g $\dots$}] ({\it istr/ibaf graph only}) to reset the default istr/ibaf graph layout
                           which corresponds to the classic-graph precursor,
                           useful if it was rearranged with the other layout options below or
                           in the drag-graph.
                                                      
\item[{\bf \parbox[t]{22ex}{\raggedleft circle/spiral $\dots$\\
                            -o/O \phantom{$\dots$} } } ]                           
                           \parbox[t]{\linewidth}{enter ``{\it circle-}{\bf o}'' to show the graph
                             as a circle, or ``{\it spiral-}{\bf O}'' as a spiral.
                              Circle layout is the default for the 1d network-graph and the jump-graph.
                           Spiral layout requires at least 30 nodes to look right.}\\
                           
\item[{\bf \parbox[t]{22ex}{\raggedleft  1d/2d(tog)/3d $\dots$\\
                            -1/2/3 \phantom{$\dots$} } } ]
                           \parbox[t]{\linewidth}{ enter ``{\bf 1}'', ``{\bf 2}'' or ``{\bf 3}''
                             to show the graph in the corresponding dimension\cite[\hspace{-1ex}\footnotesize{\#10.2}]{EDD}.
                             2d or 3d is the default for the corresponding network-graph.
                             The 2d layout can be either  square or hexagonal --- key \mbox{``{\it{2d(tog)-}{\bf 2}}''} 
                              toggles in between.}\\
    
\item[{\bf rnd-r/R $\dots$}] enter ``{\it rnd-}{\bf r}'' to ``shake'' the layout,
                             repositioning nodes randomly nearby their current 
                             position.  Enter  ``{\it rnd-}{\bf R}'' for a completely random layout.

\item[{\bf quit-q} $\dots$]  to quit the initial-graph, returning to a point where the graph
                            was selected. For example,
                            the network architecture prompt\cite[\hspace{-1ex}\footnotesize{\#17.1}]{EDD}
                            for the network-graph,
                            and the ``Attractor basin complete prompt''\cite[\hspace{-1ex}\footnotesize{\#30.4}]{EDD}
                            for the istr/ibaf graph and \mbox{f-jump-graph}, where {\bf return}
                            would generate the next mutant.
                            
\end{list}

\begin{figure}[H]
   \begin{center}
     \fbox{\includegraphics[width=.9\linewidth,viewport=0 0 1296 942, clip]{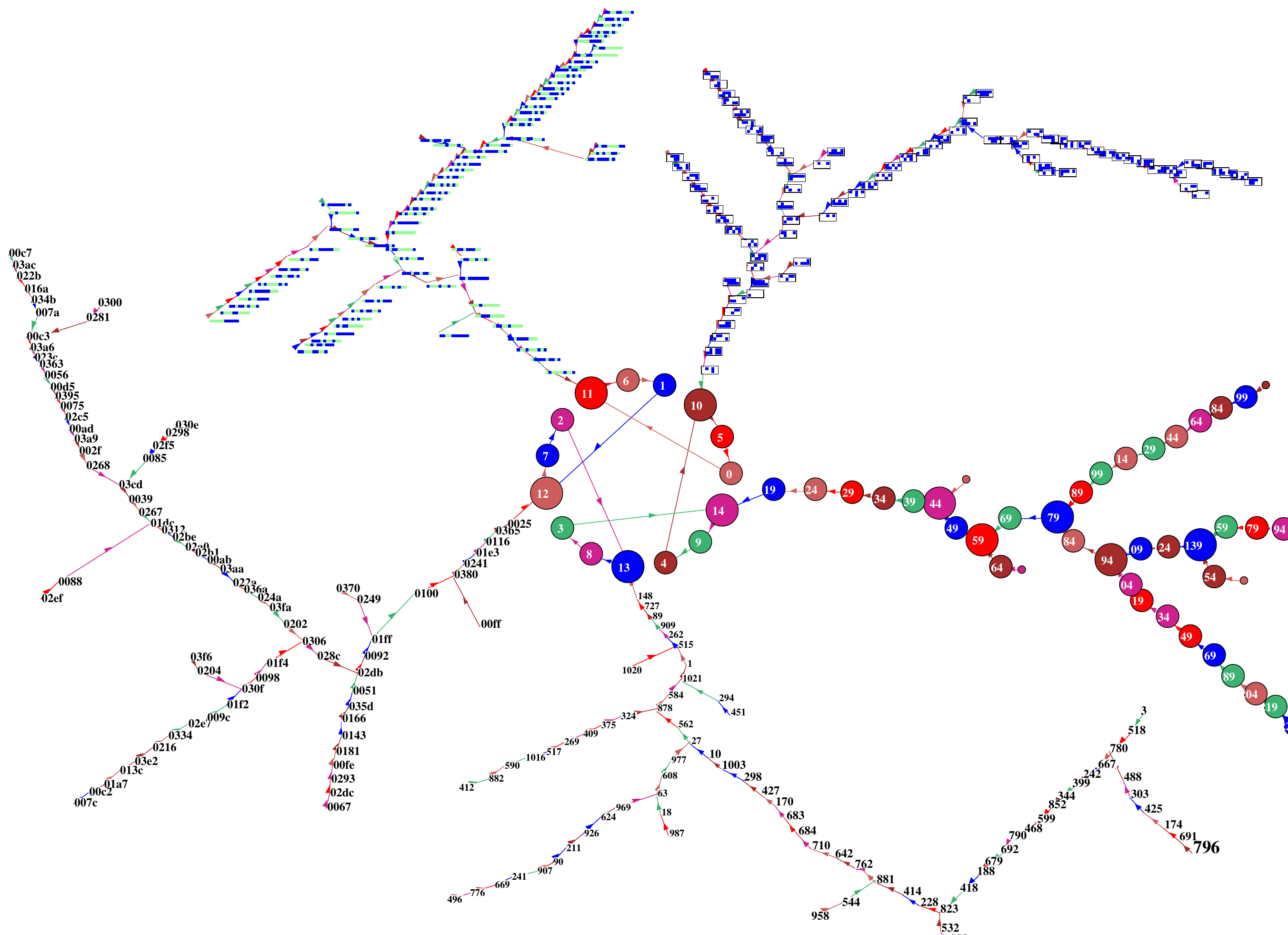}}
   \end{center} 
    \vspace{-4ex}   
    \caption [istr-graph single basin]
    {\textsf{
    This istr-graph example shows a single basin in drag mode with compression of
    equivalent subtrees, various alternative node displays, and rearranged to avoid overlap.
    Compression has reordered attractor states accounting for
    cycle cross links. This is a larger basin for the same CA as in figure~\ref{node-dis.ps}
    (1d CA $v2k3$, $n$=10, rule 30).}}
     \label{istr-graph single basin}
\end{figure}

%-----------------------------------------------------------------------%
\section{Drag-graph options}
\label{Drag-graph options}

\noindent
From the initial-graph, enter {\bf return}, or click the left or right
mouse button to switch to the \mbox{drag-graph} and its top-right
drag-reminder (section~\ref{The istr/ibaf graph drag reminder}).
Drag-graph options apply to the ``active node'',
which is either ``single'' or a node that anchors an ensemble --- the ``active fragment''.
As well as dragging, the active fragment is subject to
display/rescale/rotate/flip options
(section~\ref{Common presentation options}),
so the terms ``drag/dragging'' also imply all these actions.
Any active fragment can also be shown/printed in
isolationr\cite[\hspace{-1ex}\footnotesize{\#20.4.10}]{EDD}..

\enlargethispage{3ex}
Select the active node by a left click (and initially also a right click) with the
pointer at the center of a disc or at the tip or junction of links.
For an alphanumeric node\footnote{A drag-graph can have a mixture of node displays,
including alphanumeric
in different font sizes (figure~\ref{label146.ps}).
While dragging, all alphanumeric displays revert to the default font size
(reset in the initial-graph) to optimise animation --- once dropped
the correct font sizes are restored.}
point to the bottom-left (of the top line).
For a node shown as a 1d or 2d pattern, point to the top-left.

When activated the node is locked onto the pointer with the left button depressed 
and can be immediately dragged (with its active fragment).
The selected node's disc number appears in the title.
To drop (and deactivate) release the left button. The disc number remains in the title
until the next activation.

%Clicking empty space deactivates the node.
%If clicked and dragged with the right button, there seems to be no effect, but
%a left click makes the node (+fragment) reappear in the new position, and the node no in the title

A list of the options in the network/istr/ibaf/jump-graph drag reminder
is set out and decoded in section~\ref{drag-options decode list} in roughly the order they appear
in the reminder.
The titles in red start with \mbox{``{\color{BrickRed}{\bf node x:}}''} giving  the (most recently) active node's number
followed by the drag status, below,

\begin{list}{$\Box$}{\parsep 0ex \itemsep .6ex
  \leftmargin 15ex \rightmargin 0ex \labelwidth 20ex \labelsep 1ex}
\item[\underline{\it drag status in title} $\dots$]   \underline{\it consequences}
\item[{\color{BrickRed}{\bf single:}} $\dots$]
             to drag a single node --- enter ``{\it single-}{\bf s}'' at any time.
             Single node status allows a ``block'' (below), and also creating multi-line node
             ``labels' (figure~\ref{label146.ps}).
             Rotations and flips do not apply to a singe node.
\item[{\color{BrickRed}{\bf Block x-y:}} $\dots$]
            within single node status, enter ``{\it Block-}{\bf B}'' to define it with
            suboptions\cite[\hspace{-1ex}\footnotesize{\#20.11}]{EDD}.
            Then drag the (geometric) block (in 1d/2d/3d) noting that the active node can be
            outside a ``block'' as well as inside.
            The default is set by the last two active nodes. To leave block status
            and resume ``single node'' status, enter ``{\it exit-Block-}{\bf B}''
            or change status with one of the options ``{\bf s/i/o/e/a}''.
\item[{\bf \color{BrickRed} \parbox[t]{16ex}{\raggedleft
                inputs $\dots$ \\
                outputs \phantom{$\dots$}\\
                either \phantom{$\dots$} } } ]
             \parbox[t]{\linewidth}{enter one of the options \mbox{``{\it in/out/either-}{\bf i/o/e}''}
             at any time for the corresponding ``linked fragment'' status to appear,
             and to drag a linked fragment\cite[\hspace{-1ex}\footnotesize{\#20.4.3}]{EDD} ---
             a node and nodes linked to it by inputs, outputs, or either (meaning irrespective of direction).            
             An unlimited range ``{\color{BrickRed}{\bf step=nolimit:}}'', the default,
             appears in the title
             (or enter ``{\it nolimit-}{\bf 0}'')  ---  or set a shorter range by distance 
             measured in link-steps ``{\bf step-(1-9})'' --- for example if ``{\bf 2}'' is entered 
             ``{\color{BrickRed}{\bf step=2:}}'' appears. 
             Linked fragment status allows link cuts/additions\cite[\hspace{-1ex}\footnotesize{\#20.4.4}]{EDD}.}\\

\item[{\color{BrickRed}{\bf allnodes:}} $\dots$]
              enter ``{\bf all-a}'' at any time to drag a node and the complete graph.\\
  
\end{list}

Keyhits usually take effect without entering {\bf return}, but in some cases there
are suboptions as noted.
Options that require a given context are indicated --- for example
``\mbox{({\it linked only})}'', ``\mbox{({\it fragment  only})}'',
``\mbox{({\it single  only})}'' .
To check the type and extent of the currently defined fragment, drag any node
to see which nodes move with it.\\

\subsection{drag-options decode list}
\label{drag-options decode list}

\begin{list}{$\Box$}{\parsep 0ex \itemsep .8ex  
    \leftmargin 15ex \rightmargin 0ex \labelwidth 30ex \labelsep 1ex}
   
\item[\underline{\it options} $\dots$]   \underline{\it what they mean}
                     
\item[{\bf drag-leftb} $\dots$] click the left mouse button on a node 
                        to activate it and hold down to drag the
                        node or fragment --- then release at a new
                        position. Initially, to activate a node it may
                        be necessary to click the right button first,
                        then the left, possibly a few times.  Hence
                        the reminder:\\[1.5ex]
                        \hspace*{3ex} {\bf inactive?-rightb button-first}

\item[{\bf PScript-P} $\dots$] to save the current graph image as it appears --- a vector PostScript file
  --- suboptions appear\cite[\hspace{-1ex}\footnotesize{\#20.15}]{EDD}.

\item[{\bf elstc/snap-d} $\dots$]  to toggle between two methods 
                        of dragging (figure~\ref{snap/elastic}. By default, a node (+fragment) is dragged
                        with ``{\bf elastic}'' links and graphical animation.
                        Alternatively, animation can be suppressed and the active
                        node dragged leaving a trail --- on release, the node (+fragment)
                        and links ``{\bf snap}'' into place and the trail
                        disappears.
                        For large graphs ``{\bf snap}'' is a more efficient method.
                        Enter ``{\bf d}'' to toggle between the two --- the order of
                        ``{\bf elstc/snap}'' and ``{\bf snap/elstc}'' shows which is active.

\item[{\bf gap-g} $\dots$] ({\it linked only for the whole graph in any status})
                to recursively disconnect nodes which have only inputs or only outputs,
                \mbox{``gap-nodes''\cite[\hspace{-1ex}\footnotesize{\#20.4.9}]{EDD}},
                        so cannot transmit information.
                        Applies to the whole graph in any drag status, and can be repeated
                        until only cycles remain. Enter ``{\it net-}{\bf \#}'' to restore the original.

\item[{\bf rotate-x/X} $\dots$] ({\it fragment only})                         
                        to rotate the active fragment about any pointer position
                              by the default angle of 15 degrees. The default can be revised in the
                              initial-graph with {\bf settings-S}.
                              Enter ``{\bf x}'' to rotate clockwise, or ``{\bf X}'' anti-clockwise.

\begin{figure}[H]
  \begin{center}
   \begin{minipage}[t]{.85\linewidth}
   \begin{minipage}[t]{.5\linewidth}
      \includegraphics[height=.75\linewidth]{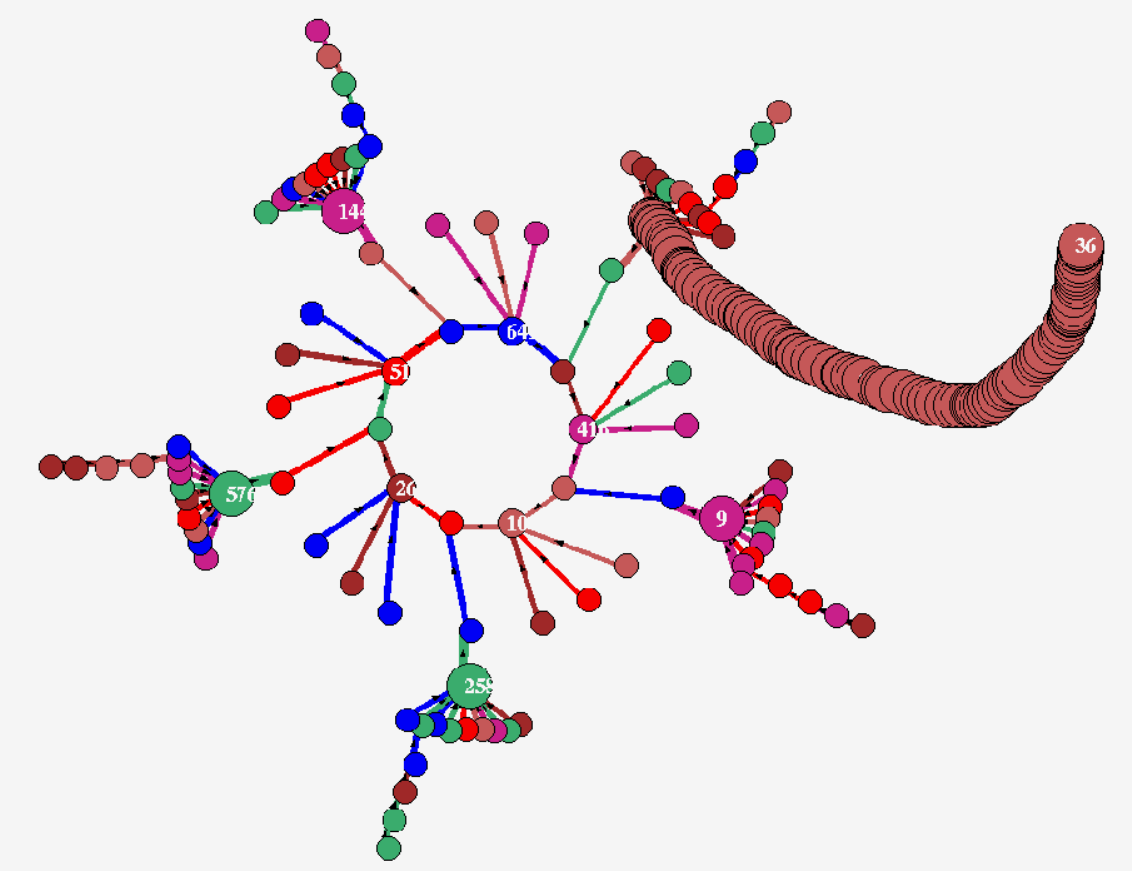}
         \vspace{-2ex} \textsf{\small (a) dragging with ``snap'' leaves a trail}
   \end{minipage}      
   %\hfill   
   \begin{minipage}[t]{.5\linewidth}
         \includegraphics[height=.75\linewidth,viewport=138 63 1152 810,clip]{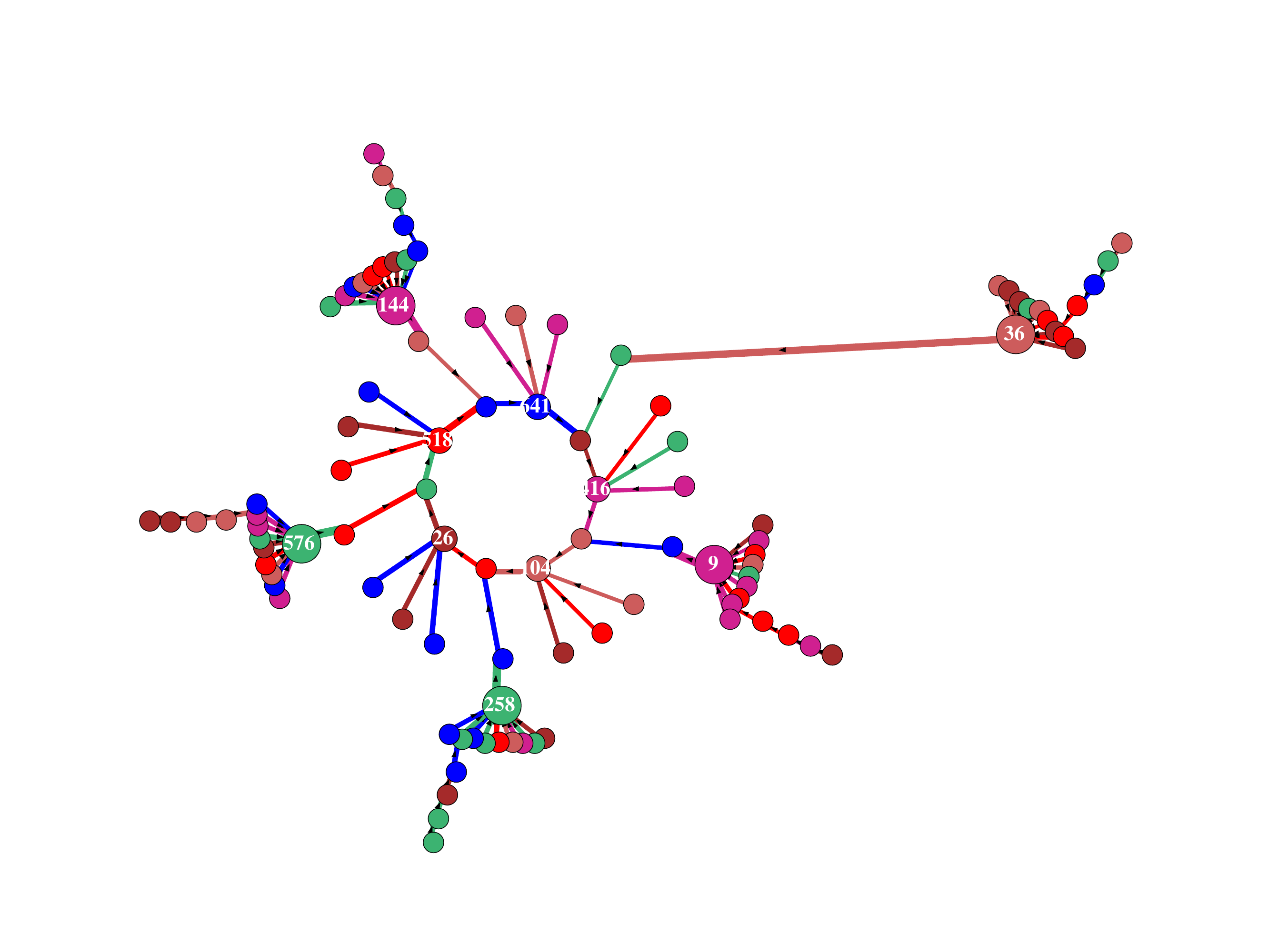}
         \vspace{-2ex} \textsf{\small (b) on release the new layout snaps into place}
   \end{minipage}
   \end{minipage}\\[-0.5ex]
   \end{center}
    \caption [dragging with ``snap'']
    {\textsf{Dragging with ``snap'' (instead of ``elastic'' --- toggle with ``{\bf d}'').
    The ibaf-graph of an isolated basin of attraction (for a 1d CA $v2k3$, $n$=10, rule 9).
    The same applies to the istr-graph but disc numbers will differ.
    (a) the default layout --- dragging node 38 to a new position
    with ``snap'' active, and the left mouse button depressed.
    (b) On left button release, node 38 and its linked fragment by
    inputs, snaps into place. This is an alternative to dragging with ``elastic'' animation.
     \label{snap/elastic}}}
\end{figure} 

\begin{figure}[H]
%\vspace{-1ex}
\begin{center}
   \includegraphics[width=.8\linewidth,viewport=150 150 1157 749,clip]{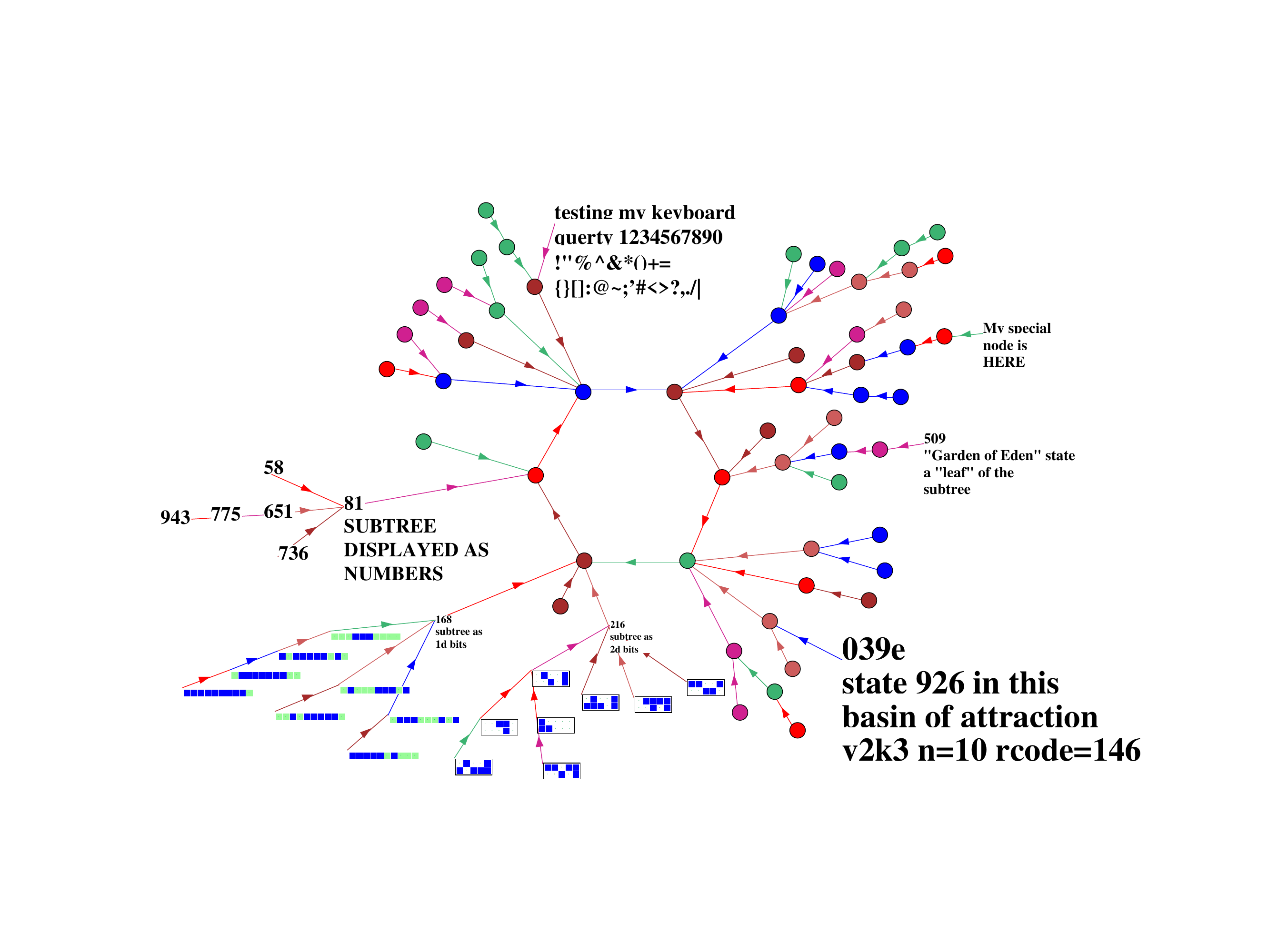}
\end{center} 
\vspace{-2ex}     
        \caption [Example of arbitrary labels]       
                 {\textsf{An example of arbitrary labels'\cite[\hspace{-1ex}\footnotesize{\#20.12}]{EDD} 
                 in an ibaf-graph with ``{\it Label}-{\bf L}'', in various text sizes.
                 The methods apply equally to the istr-graph, network-graph and jump-graph.
                 In this example the ibaf-graph (v2k3 n=10 rcode 146) has unscaled disks and edges,
                 and the basin was isolated with ``{\it just}-{\bf j}''.
                 The default label starts over its node, but can be lowered with leading line breaks
                 or displaced horizontally with spaces so as not to hide the current
                 node display, which can still be toggled (key '{\bf =}')
                 for single states or fragments.
                 \label{label146.ps}}}                 
\end{figure}

\item[{\bf flip-h/v} $\dots$] ({\it fragment only})
                         to flip the active fragment about any pointer position position,
                         enter `{\bf h}'' to flip horizontally, or ``{\bf v}'' to flip vertically.

\item[{\bf nodes-= $\dots$}] (equals sign) to cycle (toggle) though successive
                             node displays (figure~\ref{node-dis.ps}) for 
                             a single node or fragment. The toggle sequence depends
                             on the graph type as in section~\ref{Common presentation options}.

\item[{\bf nodes-E $\dots$}] ({\it fragment status only}) 
                      to equalise the display/size of all nodes in the fragment according to the active node,
                      which can be outside a ``block'' as well as inside.
                  
\item[{\bf nodes-(/) $\dots$}]  (simple brackets) enter 
                              ``{\bf (}'' to contract, or ``{\bf )}'' to expand  the current node display
                              for a single node or fragment, 
                              by default factors in the initial graph.
                             
\item[{\bf links-\{/\}} $\dots$] (curly brackets) enter
                     ``{\bf \{}'' to contract, or ``{\bf \}}'' to
                             expand the length of links (or distances between nodes for unlinked graphs)
                              for a fragment, by default factors in the initial graph.
                             
\item[{\bf both-[/]} $\dots$] (square brackets) enter
                             enter ``{\bf [}''to contract, or ``{\bf ]}''  to expand,
                             both nodes and links together, i.e. both ``{\bf nodes-(/)}''
                             and ``{\bf links-\{/\}}'' at the same time.
                             
\item[{\bf just-j/J} $\dots$] to isolate a component or fragment. ``{\bf j}'' will toggle
                     showing just a component with the active node,
                     for example a single basin in the istr/ibaf graph to work on
                     in isolation with all drag functions.
                     ``{\bf J}'' (capital) will toggle showing just the active fragment
                     linked to the active node.
                          
\item[{\bf block-B} $\dots$] ({\it ``single'' status only}) to define a block ---
                     suboptions in \cite[\hspace{-1ex}\footnotesize{\#20.11}]{EDD}.
                        The outer edges of the block can be set, or defaults accepted
                        --- the last two nodes that were activated.
                        For a linked network-graph in 2d or 3d the outer edges can be the
                         outer corners of a 2d or 3d block ---
                        {\color{BrickRed}{\bf Block 6-21}} (for example) appears in the title.
                        
\item[{\bf exit-block-B} $\dots$] ({\it in block status}) enter ``{\it exit-block-}{\bf B}''
                         to exit a block and revert to ``single'' status, or exit
                         by changing status with ``{\it single-}{\bf s}'', ``{\it in/out/either-}{\bf i/o/e}'',
                         or ``{\it all-}{\bf a}''.
                         
\item[{\bf Label-L/+} $\dots$] ({\it ``single'' status only}) enter ``{\it Label-}{\bf L}''
                         to create (or remove) a multi-line label at the active node (figure~\ref{label146.ps})
                        --- suboptions apply\cite[\hspace{-1ex}\footnotesize{\#20.12}]{EDD}.
                        Enter ``{\it Label-}{\bf +}'' (plus sign) to toggle between all current labels and the
                        current node display. Labels persist if reverting
                        to the initial-graph where all labels can also be toggled  with ``{\it Label-}{\bf +}''.
                        
\item[{\bf \parbox[t]{18ex}{\raggedleft  Lnk23: $\dots$\\
                            cut/restore \phantom{$\dots$}\\
                            -c/r \phantom{$\dots$} } } ]
                        \parbox[t]{\linewidth}{({\it linked fragment only, node 23 for example})
                        enter ``{\it cut-}{\bf c}'' to cut (disconnect)
                        the active node from its immediate links, depending on
                        which option, {\color{BrickRed}{\bf inputs}}, {\color{BrickRed}{\bf outputs}}
                        or {\color{BrickRed}{\bf either}}
                        is active. Enter ``{\it restore-}{\bf r} to undo the cuts.}\\

\item[{\bf \parbox[t]{22ex}{\raggedleft  Lnk23-19: $\dots$\\
                            cut/add/restore \phantom{$\dots$}\\
                            -C/A/R \phantom{$\dots$} } } ]
                        \parbox[t]{\linewidth}{({\it linked fragment only, nodes 23 and 19 for example})
                        The active and previously active node numbers appear 
                        in the prompt. Enter ``{\it cut-}{\bf C}'' to cut, ``{\it add-}{\bf A}'' to add
                        a link (or links) between the pairs, or ``{\it restore-}{\bf R}'' to restore 
                        --- depending on which option, {\color{BrickRed}{\bf inputs}},
                        {\color{BrickRed}{\bf outputs}}
                        or {\color{BrickRed}{\bf either}} is active.}\\

\item[{\bf net-\#} $\dots$] to restore the original links in the graph, undoing any cuts or additions.
                        Current cuts/additions are conserved when reverting to the initial-graph,
                        but can be restored with the initial option ``{\bf net-\#}''. 

\item[{\bf step-(1-9)} $\dots$] enter a number between {\bf 1} and {\bf 9} to
                        limit a linked fragment by the distance from the active node,
                        measured in link-steps (time-steps for the istr/ibaf graph) ---
                        continuous chains of directed edges
                        depending on which option, {\color{BrickRed}{\bf inputs}},
                        {\color{BrickRed}{\bf outputs}}
                        or {\color{BrickRed}{\bf either}}, is active.                        
                        For example, enter ``{\bf 1}'' for immediate links,
                        ``{\bf 2}'' up to 2 steps away, etc., up to 9 steps.
                        The current status is shown in the 
                        drag reminder, for example {\color{BrickRed}{\bf step=1}}.
                        For unlimited enter ``0'' (zero)
                        as below. This can be set in any drag status but applies for a linked fragment.
                        
\item[{\bf nolimit-0} $\dots$] enter ``{\it nolimit-}{\bf 0}'' (zero) for an unlimited linked fragment,
                        continuous chains of links relative to the active node,
                        depending on which option, {\color{BrickRed}{\bf inputs}},
                        {\color{BrickRed}{\bf outputs}}
                        or {\color{BrickRed}{\bf either}}, is active.
                        {\color{BrickRed}{\bf step=nolimit}} is shown in the drag reminder.
                        This can be set in any drag status but applies for a linked fragment.
                        
\item[{\bf single-s} $\dots$]  to set ``single'' drag status for dragging just the active node.
                        The title changes to  {\color{BrickRed}{\bf single node 23:}}
                        (for example). Single status also allows ``blocks'' and ``labels''.

\item[{\bf \parbox[t]{22ex}{\raggedleft
                in/out/either $\dots$ \\
                -i/o/e \phantom{$\dots$} } } ]                        
                         \parbox[t]{\linewidth}{enter ``{\it in-}{\bf i}'', ``{\it out-}{\bf o}''
                         or ``{\it either-}{\bf e}''
                         to define the type of link, \mbox{{\color{BrickRed}{\bf inputs}}},
                         {\color{BrickRed}{\bf outputs}}
                         or {\color{BrickRed}{\bf either}} (shown in the title) for dragging linked fragments.}\\
                         
\item[{\bf all-a} $\dots$] enter ``{\it all-}{\bf a}'' to drag all nodes, the whole graph. 
                          The title changes to {\color{BrickRed}{\bf all nodes:}}. 
\item[{\bf exit-q} $\dots$]  Enter ``{\bf q}'' to exit the drag-graph and return to
                          initial-graph and reminder (section~\ref{Initial-graph options}). Its easy to flip
                          between the two reminders to implement alternative functions. 
\end{list}

\section{Concluding remarks}
\label{Concluding remarks}

This article has presented a new generic method to render interactive
any classic attractor basin (state transition) graph in DDLab.  The
``istr-graph'' can be invoked for any type of classic-graph including
a subtree, single basin or basin of attraction field, with
compression for 1d CA. The \mbox{istr-graph} reproduces the classic-graph
layout at any stage of drawing, either complete or paused, and
respecting any prior classic settings.  This is an important and necessary
improvement on the ibaf-graph\cite{wuensche2024}
specific for the complete basin of attraction field, though the
ibaf-graph is nevertheless retained for some of its unique
attributes.

The istr-graph performs all the functions of DDLab's pre-existing
interactive graph types; the ibaf-graph released in 2024, and the long
standing network-graph and jump-graph released in
2002\cite{DDLab2002update}, with some refinements added with this update.

The most significant interactive idea for
these directed graphs is the ``fragment'' anchored to an active node
captured by the pointer, and defined by the reach of directed links
based on just inputs or just outputs, or links in either direction,
and an unlimited or restricted distance in link steps, which are
time-steps for istr/ibaf graphs representing dissipative dynamical
systems.  The fragment is dragged/dropped as the active node is moved
by the pointer, and also serves as the node ensemble for relabelling,
dilating, manipulating and isolating the fragment, so permits
analysis and deconstruction of the graph with immediate and continuous
feedback.

The descriptions and instructions of the various functions, and the
snap-shot illustrations, give just a flavour of the istr-graph and the
other interactive visualisations in DDLab.
Hands on playing around experience and experiment is recommended to
get to know the scope and possibilities, in conjunction with the
latest update of the in depth online reference manual ``Exploring
Discrete Dynamics''\cite{EDD}, especially chapter 20.

Compiled versions, and the code itself in the c language are available
under the GNU General Public License for various platforms and
operating systems\cite{Wuensche-DDLab} including Linux and Mac.

\end{document}